\documentclass[12pt]{article}
\usepackage{amssymb}
\usepackage{amsmath}
\usepackage{graphicx}
\usepackage{indentfirst}
\usepackage{cite}

\allowdisplaybreaks

\begin{document}

\title{Cross sections for inelastic meson-meson scattering}
\author{Kai Yang$^1$, Xiao-Ming Xu$^1$, and H. J. Weber$^2$}
\date{}
\maketitle \vspace{-1cm}
\centerline{$^1$Department of Physics, Shanghai University, Baoshan,
Shanghai 200444, China}
\centerline{$^2$Department of Physics, University of Virginia, Charlottesville,
VA 22904, USA}

\begin{abstract}
We study two kinds of inelastic meson-meson scattering. The first kind is 
inelastic 2-to-2 meson-meson scattering that is governed by quark
interchange as well as quark-antiquark annihilation and creation. Cross-section
formulas are provided to get unpolarized cross sections for
$\pi K \to \rho K^\ast$ for $I=1/2$, $\pi K^\ast \to \rho K$ for $I=1/2$,
$\pi K^\ast \to \rho K^\ast$ for $I=1/2$, and
$\rho K \to \rho K^\ast$ for $I=1/2$. Near threshold, quark
interchange dominates the reactions near the critical temperature. The second
kind is 2-to-1 meson-meson scattering with the process that
a quark in an initial meson and an antiquark in another initial meson
annihilate into a gluon and subsequently the gluon is absorbed by the other
antiquark or quark. The transition potential for the process is derived. Four
Feynman diagrams at tree level contribute to the 2-to-1
meson-meson scattering. Starting from the $S$-matrix element,
the isospin-averaged unpolarized cross section with transition amplitudes
is derived. The
cross sections for $\pi \pi \to \rho$ and $\pi K \to K^*$ decrease with
increasing temperature.
\end{abstract}



\vspace{0.5cm}
\leftline{\bf I. INTRODUCTION}
\vspace{0.5cm}

Meson-meson scattering is an important field in exploring strong interactions.
Starting from the chiral perturbation theory Lagrangian, any amplitude for
meson-meson scattering is derived to form a perturbative expansion in powers of
external momenta and quark masses \cite{GL,GMei,BCEGS,GO,NP}. Depending on what
mesons scatter, the calculations of the amplitudes have reached the fourth or 
sixth power of the external momenta. This expansion is successful 
in the description of low-energy meson-meson scattering, but cannot
describe resonances. In extending chiral perturbation theory to the study of
meson-meson scattering beyond the low-energy regime, nonperturbative schemes,
for example, the inverse amplitude method, have been proposed. Consequently, 
resonances can be reproduced and studied within the nonperturbative schemes 
together with the chiral expansion.
Since the elastic phase shifts for $\pi \pi$ scattering and $\pi K$ scattering
can be measured, resonances that contribute to the phase shifts have been 
studied in the Roy 
equations~\cite{Roy}, the Pad\'e method~\cite{DHT}, the inverse amplitude 
method~\cite{GO,NP,Hannah}, the large-$N_f$ expansion~\cite{DP}, 
the $K$-matrix method~\cite{ZB}, the master formula approach~\cite{SYZ}, the
current algebra unitarization~\cite{BBO}, the coupled-channel unitary 
approaches~\cite{OO}, the $N/D$ method~\cite{CM}, the Bethe-Salpeter
approach~\cite{NA}, and the approaches based on effective meson
Lagrangians~\cite{JPHS,Li,BFS}. Nevertheless, some isospin channels of the 
reactions do not involve resonances, and are instead
governed, for example, by a quark-interchange process or by annihilation of
a quark-antiquark pair and subsequent creation of another quark-antiquark
pair. The elastic meson-meson scattering governed by quark interchange is,
for example, $\pi \pi$ for $I=2$ and $\pi K$ for $I=3/2$, which have been
studied in Ref. \cite{BS} in the quark interchange mechanism. The inelastic
meson-meson scattering governed by quark interchange is \cite{LX,SX1}, for
example, 
$\pi\pi \to \rho\rho$ for $I=2$, $KK \to K^* K^*$ for $I=1$, $KK^* \to K^*K^*$ 
for $I=1$, $\pi K \to \rho K^*$ for $I=3/2$, $\pi K^* \to \rho K^*$ for 
$I=3/2$, $\rho K \to \rho K^*$ for $I=3/2$, and $\pi K^* \to \rho K$ for 
$I=3/2$. Cross sections for these channels of endothermic reactions 
have the characteristic that the cross
sections rise very rapidly from threshold energies, arrive at maximum values,
and decrease rapidly. The inelastic meson-meson scattering governed by
quark-antiquark annihilation and creation is \cite{SXW}, for example,
$\pi \pi \to \rho \rho$ for $I=1$, $\pi \pi \to K \bar K$ for $I=1$, 
$\pi \rho \to K \bar {K}^\ast$, $\pi \rho \to K^* \bar{K}$,
$K \bar {K} \to K^* \bar {K}^\ast$, and
$K \bar{K}^\ast \to K^* \bar{K}^\ast$.
The cross sections for the endothermic reactions
have the characteristic that
they may decrease from maximum values very slowly.
In Refs. \cite{LDHS,BKWX} isospin-averaged cross sections for 
$\pi \pi \to K \bar K$, $\rho \rho \to K \bar K$, $\pi \rho \to K \bar {K}^*$, 
and $\pi \rho \to K^* \bar {K}$ have been obtained from effective meson 
Lagrangians via the exchange of either a
kaon or a vector kaon between the two initial mesons. 
The reactions governed by quark interchange or by quark-antiquark
annihilation and creation can also be studied by implementing chiral
perturbation theory within the nonperturbative schemes. For example, elastic 
$\pi \pi$ scattering for $I=2$ and elastic $\pi K$ scattering for $I=3/2$ 
have been stuided in Refs. \cite{GO,NP,Roy,DHT,Hannah,OO}. The Lagrangian of
chiral perturbation theory includes various couplings of pseudoscalar mesons,
which establish amplitudes for the scattering.
The experimental data of $S$-wave elastic phase shifts for the scattering are 
reproduced from the amplitudes by means of the nonperturbative schemes.

In the present work
we first study inelastic meson-meson scattering mediated by both quark 
interchange and quark-antiquark annihilation into a gluon which
subsequently creates a quark-antiquark pair. This kind
of scattering includes $\pi K \to \rho K^*$ for $I=1/2$, 
$\pi K^* \to \rho K$ for $I=1/2$,
$\pi K^* \to \rho K^*$ for $I=1/2$, and $\rho K \to \rho {K}^*$ for $I=1/2$.
Since the four isospin channels of these reactions have not been taken into 
account in models for ultrarelativistic heavy-ion collisions,
our results will be helpful in improving these models.
Second, we study the reactions $\pi \pi \to \rho$ and $\pi K \to K^*$ based on
quark-antiquark annihilation into a gluon which is further absorbed by a
quark or an antiquark. The cross sections we obtain will be compared to 
experimental data on $\pi \pi \to \rho$ \cite{FMMR} and the estimate derived
in Ref. \cite{Ko} on $\pi K \to K^*$ in vacuum. Since both reactions are
important in hadronic matter, we also study their temperature dependence.

Taking the parameter in the Feynman-Hellmann theorem \cite{Hellmann} as the
quark mass, the theorem can be used to study the dependence of 
the ground-state hadron mass on the quark mass. The theorem has been 
recently generalized to unstable states in quantum field theory \cite{EMRS}.
Combining the theorem and recent lattice data on the quark-mass dependence 
of hadron masses, Elvira et al. \cite{EMRS} have studied the possible exotic
admixture (pentaquarks, tetraquarks, etc.) of hadronic states.
They conclude that the ground-state vector 
mesons are predominantly quark-antiquark states. We thus assume that the 
ground-state vector mesons are quark-antiquark states.

While Barnes et al. \cite{BS} study elastic $\pi \pi$ scattering for $I=2$
and elastic $\pi K$ scattering for $I=3/2$, they assume that Born-order
quark-interchange diagrams dominate the elastic scattering and 
that $s$-channel resonance production and $t$-channel resonance exchange
are not important. While we study inelastic meson-meson scattering in Ref.
\cite{SXW}, we assume that diagrams of Born-order quark-antiquark annihilation 
and creation dominate the inelastic scattering and that the contributions
of the annihilation with gluonic excitation into a hybrid meson and
the annihilation of a color-singlet quark-antiquark pair into a (virtual) 
glueball \cite{WI} are negligible. In the present work we assume that the
diagrams of Born-order quark interchange and the diagrams of Born-order 
quark-antiquark annihilation and creation dominate the four reactions: 
$\pi K \to \rho K^*$ for $I=1/2$, $\pi K^* \to \rho K$ for $I=1/2$,
$\pi K^* \to \rho K^*$ for $I=1/2$, and $\rho K \to \rho {K}^*$ for $I=1/2$.

This paper is organized as follows. In Sect.~II we provide cross section 
formulas for 2-to-2 meson-meson reactions that are governed not only by quark
interchange but also by quark-antiquark
annihilation and creation. In Sect.~III we derive the formulas of 
isospin-averaged unpolarized cross sections for 2-to-1 meson-meson reactions.
In Sect.~IV we derive a transition potential for the process that is a pair of
quark-antiquark annihilates into a gluon and subsequently the gluon is
absorbed by a quark or an antiquark. Transition amplitudes in the cross section
formulas are calculated. In Sect.~V numerical cross sections are presented,
and relevant discussions are given. In Sect.~VI we summarize the present work.

\vspace{0.5cm}
\leftline{\bf II. CROSS-SECTION FORMULAS FOR 2-TO-2 REACTIONS}
\vspace{0.5cm}

Quark interchange leads to the meson-meson scattering processes 
$A(q_1\bar{q}_1)+B(q_2\bar{q}_2) \to C(q_1\bar{q}_2)+D(q_2\bar{q}_1)$.
The scattering may take the prior form where gluon exchange 
takes place prior to quark interchange or the post form where quark interchange
is followed by gluon exchange. The transition amplitude corresponding to
the prior form is
\begin{eqnarray}
{\cal M}_{\rm fi}^{\rm prior} & = &
\sqrt{2E_A2E_B2E_C2E_D}\int \frac{d^3 p_{q_1\bar{q}_2}}{({2\pi})^3} 
\frac{d^3 p_{q_2\bar{q}_1}}{({2\pi})^3}
\psi_{q_1\bar{q}_2}^+(\vec{p}_{q_1\bar{q}_2})
\psi_{q_2\bar{q}_1}^+(\vec{p}_{q_2\bar{q}_1})
      \nonumber  \\
& &
(V_{q_1\bar{q}_2}+V_{\bar{q}_1 q_2}+V_{q_1 q_2}+V_{\bar{q}_1\bar{q}_2})
\psi_{q_1\bar{q}_1}(\vec{p}_{q_1\bar{q}_1})
\psi_{q_2\bar{q}_2}(\vec{p}_{q_2\bar{q}_2}) ,
\end{eqnarray}
and the one corresponding to the post form is
\begin{eqnarray}
{\cal M}_{\rm fi}^{\rm post} & = &
\sqrt{2E_A2E_B2E_C2E_D}\int \frac{d^3 p_{q_1\bar{q}_1}}{({2\pi})^3} 
\frac{d^3 p_{q_2\bar{q}_2}}{({2\pi})^3}
\psi_{q_1\bar{q}_2}^+(\vec{p}_{q_1\bar{q}_2})
\psi_{q_2\bar{q}_1}^+(\vec{p}_{q_2\bar{q}_1})
      \nonumber  \\
& &
(V_{q_1\bar{q}_1}+V_{\bar{q}_2 q_2}+V_{q_1 q_2}+V_{\bar{q}_1\bar{q}_2})
\psi_{q_1\bar{q}_1}(\vec{p}_{q_1\bar{q}_1})
\psi_{q_2\bar{q}_2}(\vec{p}_{q_2\bar{q}_2}) ,
\end{eqnarray}
where $\vec{p}_{ab}$ is the relative momentum of constituents $a$ and $b$;
$E_A$ ($E_B$, $E_C$, $E_D$) is the energy of meson $A$ ($B$, $C$, $D$);
$\psi_{ab}(\vec{p}_{ab})$ is the mesonic quark-antiquark wave function which
is the product of the color wave function, the spin wave function, the flavor
wave function, and the relative-motion wave function of constituents $a$ and
$b$; $\psi_{ab}^+$ is the Hermitean conjugate of $\psi_{ab}$;
$V_{ab}$ is the quark potential \cite{JSX} that is given by perturbative
QCD with loop corrections to one-gluon exchange at short distances, that
becomes a distance-independent and temperature-dependent potential at long
distances, and that has a spin-spin interaction with relativistic 
modifications. The Fourier transform of the relative-motion part of 
$\psi_{ab}(\vec{p}_{ab})$ is a solution of the Schr\"odinger equation with
the potential. The meson masses obtained from the Schr\"odinger equation with
the potential at zero temperature are close to
the experimental masses of $\pi$, $\rho$, $K$, $K^*$,
$J/\psi$, $\psi'$, $\chi_{c}$, $D$, $D^*$, $D_s$, and $D^*_s$ 
mesons~\cite{PDG}. Moreover, the experimental data of $S$-wave phase shifts
for the elastic $\pi \pi$ scattering for $I=2$ in 
vacuum~\cite{pipiqi} are reproduced in the Born approximation \cite{JSX}.
In reproducing the experimental meson masses the masses of the up quark, the
down quark, the strange quark, and the charm quark are 0.32 GeV, 0.32 GeV,
0.5 GeV, and 1.51 GeV, respectively.

If a quark and an antiquark annihilate into a gluon, this gluon may create a
quark-antiquark pair. This process of quark-antiquark annihilation
and creation leads to
$A(q_1\bar{q}_1)+B(q_2\bar{q}_2) \to C(q_3\bar{q}_1)+D(q_2\bar{q}_4)$ and
$A(q_1\bar{q}_1)+B(q_2\bar{q}_2) \to C(q_1\bar{q}_4)+D(q_3\bar{q}_2)$. The
transtion amplitude for
$A(q_1\bar{q}_1)+B(q_2\bar{q}_2) \to C(q_3\bar{q}_1)+D(q_2\bar{q}_4)$ is
\begin{eqnarray}
{\cal M}_{{\rm a}q_1\bar {q}_2} & = &
\sqrt {2E_A2E_B2E_C2E_D}
\int \frac {d^3p_{q_1\bar{q}_1}}{(2\pi)^3}\frac {d^3p_{q_2\bar{q}_2}}{(2\pi)^3}
                       \nonumber         \\
& &
\psi^+_{q_3\bar {q}_1} (\vec {p}_{q_3\bar {q}_1})
\psi^+_{q_2\bar {q}_4} (\vec {p}_{q_2\bar {q}_4})
V_{{\rm a}q_1\bar{q}_2}(\vec{k})
\psi_{q_1\bar {q}_1} (\vec {p}_{q_1\bar {q}_1})
\psi_{q_2\bar {q}_2} (\vec {p}_{q_2\bar {q}_2}),
\end{eqnarray}
and the transition amplitude for
$A(q_1\bar{q}_1)+B(q_2\bar{q}_2) \to C(q_1\bar{q}_4)+D(q_3\bar{q}_2)$ is
\begin{eqnarray}
{\cal M}_{{\rm a}\bar{q}_1q_2} & = &
\sqrt {2E_A2E_B2E_C2E_D}
\int \frac {d^3p_{q_1\bar{q}_1}}{(2\pi)^3}\frac {d^3p_{q_2\bar{q}_2}}{(2\pi)^3}
                       \nonumber         \\
& &
\psi^+_{q_1\bar {q}_4} (\vec {p}_{q_1\bar {q}_4})
\psi^+_{q_3\bar {q}_2} (\vec {p}_{q_3\bar {q}_2})
V_{{\rm a}\bar{q}_1q_2}(\vec{k})
\psi_{q_1\bar {q}_1} (\vec {p}_{q_1\bar {q}_1})
\psi_{q_2\bar {q}_2} (\vec {p}_{q_2\bar {q}_2}),
\end{eqnarray}
where $\vec k$ is the gluon momentum, and
$V_{{\rm a}q_1\bar{q}_2}$ and $V_{{\rm a}\bar{q}_1q_2}$ are the transition
potentials for $q_1+\bar{q}_2 \to q_3+\bar{q}_4$ and
$\bar{q}_1+q_2 \to q_3+\bar{q}_4$, respectively. The transition potentials
have been given in Ref. \cite{SXW}. With the transition amplitudes
the experimental data~\cite{pipianni}
of $S$-wave $I=0$ and $P$-wave $I=1$ elastic phase shifts for $\pi \pi$
scattering near the threshold energy in vacuum are reproduced.

Meson $i$ ($i=A$, $B$, $C$, $D$) has the angular momentum $J_i$ with its
magnetic projection quantum number $J_{iz}$. In the center-of-mass frame
meson $A$ has the momentum $\vec P$, meson $C$ has the momentum $\vec{P}'$,
and the angle between $\vec P$ and $\vec{P}'$ is $\theta$. In hadronic matter
the cross section depends on temperature $T$ and $s=(P_A+P_B)^2$, where
$P_A$ and $P_B$ are the four-momenta of mesons $A$ and $B$, respectively.
The cross section for the scattering including the quark-interchange process 
in the prior form as well as the quark-antiquark annihilation and creation is
\begin{eqnarray}
\sigma^{\rm prior}_{\rm unpol} (\sqrt {s},T) 
& = & \frac {1}{(2J_A+1)(2J_B+1)}
\frac{1}{32\pi s}\frac{|\vec{P}^{\prime }(\sqrt{s})|}{|\vec{P}(\sqrt{s})|}
              \nonumber    \\
& & \int_{0}^{\pi }d\theta \sum\limits_{J_{Az}J_{Bz}J_{Cz}J_{Dz}}
\mid {\cal M}_{{\rm a}q_1\bar{q}_2}+{\cal M}_{{\rm a}\bar{q}_1q_2}
+ {\cal M}_{\rm fi}^{\rm prior} \mid^2
\sin \theta .
\end{eqnarray}
The cross section for the scattering including the quark-interchange process 
in the post form as well as the quark-antiquark annihilation and creation is
\begin{eqnarray}
\sigma^{\rm post}_{\rm unpol} (\sqrt {s},T) 
& = & \frac {1}{(2J_A+1)(2J_B+1)}
\frac{1}{32\pi s}\frac{|\vec{P}^{\prime }(\sqrt{s})|}{|\vec{P}(\sqrt{s})|}
              \nonumber    \\
& & \int_{0}^{\pi }d\theta \sum\limits_{J_{Az}J_{Bz}J_{Cz}J_{Dz}}
\mid {\cal M}_{{\rm a}q_1\bar{q}_2}+{\cal M}_{{\rm a}\bar{q}_1q_2}
+ {\cal M}_{\rm fi}^{\rm post} \mid^2
\sin \theta .
\end{eqnarray}
The unpolarized cross section is given by
\begin{equation}
\sigma^{\rm unpol}(\sqrt {s},T) = \frac {1}{2}
[\sigma^{\rm prior}_{\rm unpol}(\sqrt {s},T)
+\sigma^{\rm post}_{\rm unpol}(\sqrt {s},T)].
\end{equation}

\vspace{0.5cm}
\leftline{\bf III. CROSS-SECTION FORMULAS FOR 2-TO-1 REACTIONS}
\vspace{0.5cm}

If a quark in meson $A$ and an antiquark in meson $B$
annihilate into a gluon, this gluon may be absorbed by the antiquark in meson 
$A$ or the quark in meson $B$. We have four Feynman diagrams shown in 
Fig. 1 for the reaction $A(q_1\bar{q}_1)+B(q_2\bar{q}_2) \to 
H(q_2\bar{q}_1~{\rm or}~q_1\bar{q}_2)$ where meson $H$ is $q_2 \bar{q}_1$ in 
the two upper diagrams or $q_1\bar{q}_2$ in the two lower diagrams.
The $S$-matrix element for $A+B \to H$ is
\begin{eqnarray}
S_{\rm fi} & = & \delta_{\rm fi} - 2\pi i \delta (E_{\rm f} - E_{\rm i})
(<H \mid V_{{\rm r}q_1\bar {q}_2 \bar{q}_1}\mid A,B> + <H \mid V_{{\rm r}
q_1\bar {q}_2 q_2}\mid A,B> \nonumber\\
& &
+ <H \mid V_{{\rm r}q_2 \bar {q}_1 q_1}\mid A,B> + <H \mid V_{{\rm r}
q_2 \bar {q}_1 \bar{q}_2}\mid A,B>)
\end{eqnarray}
where $E_{\rm i}$ is the total energy of the two initial mesons; $E_{\rm f}$
is the energy of meson $H$; $V_{{\rm r}q_1\bar{q}_2\bar{q}_1}$ 
($V_{{\rm r}q_1\bar{q}_2q_2}$) represents the transition potential for the
annihilation of $q_1$ and $\bar{q}_2$ into a gluon and the subsequent
absorption of the
gluon by $\bar{q}_1$ in meson $A$ ($q_2$ in meson $B$);
$V_{{\rm r}q_2\bar{q}_1q_1}$ ($V_{{\rm r}q_2\bar{q}_1\bar{q}_2}$) represents
the transition potential for the annihilation of $q_2$ and $\bar{q}_1$ into 
a gluon and the subsequent absorption of the
gluon by $q_1$ in meson $A$ ($\bar{q}_2$ in meson $B$).
The wave function of mesons $A$ and $B$ is
\begin{equation}
\psi_{q_1\bar {q}_1, q_2\bar {q}_2}=
\frac {e^{i\vec {P}_{q_1\bar {q}_1}\cdot \vec {R}_{q_1\bar {q}_1}}}{\sqrt V}
\psi_{q_1\bar {q}_1} (\vec {r}_{q_1\bar {q}_1})
\frac {e^{i\vec {P}_{q_2\bar {q}_2}\cdot \vec {R}_{q_2\bar {q}_2}}}{\sqrt V}
\psi_{q_2\bar {q}_2} (\vec {r}_{q_2\bar {q}_2}),
\end{equation}
where $\vec {P}_{ab}$, $\vec {R}_{ab}$, and $\vec {r}_{ab}$ are the total
momentum, the center-of-mass coordinate, and the relative coordinate of $a$
and $b$, respectively. The wave function of meson $H$ is
\begin{equation}
\psi_{q_2\bar {q}_1}=
\frac {e^{i\vec {P}_{q_2\bar {q}_1}\cdot \vec {R}_{q_2\bar {q}_1}}}{\sqrt V}
\psi_{q_2\bar {q}_1} (\vec {r}_{q_2\bar {q}_1}),
\end{equation}
corresponding to the two upper diagrams or
\begin{equation}
\psi_{q_1\bar {q}_2}=
\frac {e^{i\vec {P}_{q_1\bar {q}_2}\cdot \vec {R}_{q_1\bar {q}_2}}}{\sqrt V}
\psi_{q_1\bar {q}_2} (\vec {r}_{q_1\bar {q}_2}),
\end{equation}
corresponding to the two lower diagrams. Every meson wave 
function is normalized in the volume $V$.

Now we derive cross section formulas for $A+B \to H$. For this we denote
by $\vec {R}_{\rm total}$ the center-of-mass coordinate of the two
initial mesons or the final meson, $\vec {P}_{\rm i}$ ($\vec {P}_{\rm f}$) the
total momentum of the two initial mesons (the final meson),
$\vec{r}_c$ the position vector of constituent $c$, and $m_c$
the mass of constituent $c$.
Meson $i$ $(i=A, B, H)$ has the mass $m_i$ and the four-momentum
$P_i=(E_{i},\vec{P}_i)$.

We first consider the two upper diagrams in Fig. 1. Three independent 
constituent position-vectors are $\vec{r}_{q_1}$, $\vec{r}_{\bar{q}_1}$, and 
$\vec{r}_{q_2}$. They are related to $\vec {R}_{q_2\bar{q}_1}$, 
$\vec{r}_{q_1\bar{q}_1}$, and $\vec{r}_{q_2\bar{q}_2}$ by
\begin{equation}
\vec{r}_{q_1} = \vec {R}_{q_2\bar{q}_1}
+ \frac {m_{\bar{q}_1}}{m_{{q}_2} + m_{\bar{q}_1}}
\vec{r}_{q_1\bar{q}_1}
- \frac {m_{{q}_2}}{m_{{q}_2}+m_{\bar{q}_1}}
\vec{r}_{q_2\bar{q}_2},
\end{equation}
\begin{equation}
\vec{r}_{\bar{q}_1} = \vec {R}_{q_2\bar{q}_1}
- \frac {m_{{q}_2}}{m_{{q}_2} + m_{\bar{q}_1}}
\vec{r}_{q_1\bar{q}_1}
- \frac {m_{{q}_2}}{m_{{q}_2}+m_{\bar{q}_1}}
\vec{r}_{q_2\bar{q}_2},
\end{equation}
\begin{equation}
\vec{r}_{q_2} = \vec {R}_{q_2\bar{q}_1}
+ \frac {m_{\bar{q}_1}}{m_{{q}_2} + m_{\bar{q}_1}}
\vec{r}_{q_1\bar{q}_1}
+ \frac {m_{\bar{q}_1}}{m_{{q}_2}+m_{\bar{q}_1}}
\vec{r}_{q_2\bar{q}_2},
\end{equation}
which lead to
\begin{equation}
d\vec{r}_{q_1} d\vec{r}_{\bar{q}_1} d\vec{r}_{q_2}=
d\vec{r}_{q_1\bar{q}_1} d\vec{r}_{q_2\bar{q}_2} d\vec{R}_{\rm total}.
\end{equation}
From the wave functions of the initial and final mesons and the independent 
position vectors we have for the left upper diagram:
\begin{eqnarray}
<H \mid V_{{\rm r}q_1\bar {q}_2 \bar{q}_1}\mid A,B> & = &
<q_2\bar{q}_1 \mid V_{{\rm r}q_1\bar{q}_2\bar{q}_1} \mid q_1\bar{q}_1, 
q_2\bar{q}_2>
    \nonumber    \\
& = & \int d\vec{r}_{q_1} d\vec{r}_{\bar{q}_1} d\vec{r}_{q_2}
\frac {e^{-i\vec{P}_{q_2\bar{q}_1}\cdot\vec{R}_{q_2\bar{q}_1}}}{\sqrt V}
\psi_{q_2\bar{q}_1}^+ (\vec{r}_{q_2\bar{q}_1})
    \nonumber    \\
& &
V_{{\rm r}q_1\bar{q}_2 \bar{q}_1}
\frac {e^{i\vec{P}_{q_1\bar{q}_1}\cdot\vec{R}_{q_1\bar{q}_1}}}{\sqrt V}
\psi_{q_1\bar{q}_1}(\vec{r}_{q_1\bar{q}_1})
\frac {e^{i\vec{P}_{q_2\bar{q}_2}\cdot\vec{R}_{q_2\bar{q}_2}}}{\sqrt V}
\psi_{q_2\bar{q}_2}(\vec{r}_{q_2\bar{q}_2})
    \nonumber    \\
& = & \int d\vec{r}_{q_1\bar{q}_1} d\vec{r}_{q_2\bar{q}_2} 
d\vec{R}_{\rm total}
V^{-\frac {3}{2}}e^{-i\vec{P}_{\rm f}\cdot\vec{R}_{\rm total}}
\psi_{q_2\bar{q}_1}^+ (\vec{r}_{q_2\bar{q}_1})
    \nonumber    \\
& &
V_{{\rm r}q_1\bar{q}_2 \bar{q}_1}\psi_{q_1\bar{q}_1} (\vec{r}_{q_1\bar{q}_1})
\psi_{q_2\bar{q}_2} (\vec{r}_{q_2\bar{q}_2})
e^{i\vec{P}_{\rm i}\cdot\vec{R}_{\rm total}
+i\vec{p}_{q_1\bar{q}_1,q_2\bar{q}_2}\cdot\vec{r}_{q_1\bar{q}_1,q_2\bar{q}_2}}
    \nonumber    \\
& = & (2\pi)^3 \delta^3 (\vec{P}_{\rm f} - \vec{P}_{\rm i})
\frac {{\cal M}_{{\rm r}q_1\bar{q}_2 \bar{q}_1}}
{V^{\frac {3}{2}}\sqrt{2E_A2E_B2E_H}},
\end{eqnarray}
where $\vec {r}_{q_1\bar {q}_1,q_2\bar {q}_2}$ 
($\vec {p}_{q_1\bar {q}_1,q_2\bar {q}_2}$) is
the relative coordinate (the relative momentum) of $q_1\bar {q}_1$
and $q_2\bar {q}_2$, and the transition amplitude is
\begin{eqnarray}
{\cal M}_{{\rm r}q_1\bar {q}_2 \bar{q}_1}
& = &
\sqrt {2E_A2E_B2E_H}
\int d\vec{r}_{q_1\bar{q}_1} d\vec{r}_{q_2\bar{q}_2}
\psi_{q_2\bar{q}_1}^+ (\vec {r}_{q_2\bar{q}_1})V_{{\rm r}q_1\bar{q}_2\bar{q}_1}
    \nonumber   \\
& &
\psi_{q_1\bar{q}_1} (\vec {r}_{q_1\bar {q}_1})
\psi_{q_2\bar{q}_2} (\vec {r}_{q_2\bar {q}_2})
e^{i\vec{p}_{q_1\bar{q}_1,q_2\bar{q}_2}
\cdot\vec{r}_{q_1\bar{q}_1,q_2\bar{q}_2}}.
\end{eqnarray}
For the right upper diagram we obtain
\begin{eqnarray}
<H \mid V_{{\rm r}q_1\bar {q}_2 q_2}\mid A,B>
& = & 
<q_2\bar{q}_1 \mid V_{{\rm r}q_1\bar{q}_2q_2} \mid q_1\bar{q}_1, q_2\bar{q}_2>
     \nonumber  \\
& = & (2\pi)^3 \delta^3(\vec{P}_{\rm f}-\vec{P}_{\rm i})
\frac{{\cal M}_{{\rm r}q_1\bar{q}_2 q_2}}{V^{\frac{3}{2}}\sqrt{2E_A2E_B2E_H}},
\end{eqnarray}
with the transition amplitude:
\begin{eqnarray}
{\cal M}_{{\rm r}q_1\bar {q}_2 q_2}
& = &
\sqrt {2E_A2E_B2E_H}
\int d\vec{r}_{q_1\bar{q}_1} d\vec{r}_{q_2\bar{q}_2}
\psi_{q_2\bar{q}_1}^+ (\vec {r}_{q_2\bar{q}_1})V_{{\rm r}q_1\bar{q}_2 q_2}
    \nonumber   \\
& &
\psi_{q_1\bar{q}_1} (\vec {r}_{q_1\bar {q}_1})
\psi_{q_2\bar{q}_2} (\vec {r}_{q_2\bar {q}_2})
e^{i\vec{p}_{q_1\bar{q}_1,q_2\bar{q}_2}
\cdot\vec{r}_{q_1\bar{q}_1,q_2\bar{q}_2}}.
\end{eqnarray}

Next, we consider the two lower diagrams in Fig. 1. The three independent
constituent position-vectors are $\vec{r}_{q_1}$, $\vec{r}_{\bar{q}_1}$, and 
$\vec{r}_{\bar{q}_2}$. They are related to $\vec {R}_{q_1\bar{q}_2}$, 
$\vec{r}_{q_1\bar{q}_1}$, and $\vec{r}_{q_2\bar{q}_2}$ by
\begin{equation}
\vec{r}_{q_1} = \vec {R}_{q_1\bar{q}_2}
+ \frac {m_{\bar{q}_2}}{m_{{q}_1} + m_{\bar{q}_2}}
\vec{r}_{q_1\bar{q}_1}
+ \frac {m_{\bar{q}_2}}{m_{{q}_1} + m_{\bar{q}_2}}
\vec{r}_{q_2\bar{q}_2},
\end{equation}
\begin{equation}
\vec{r}_{\bar{q}_1} = \vec {R}_{q_1\bar{q}_2}
- \frac {m_{{q}_1}}{m_{{q}_1} + m_{\bar{q}_2}}
\vec{r}_{q_1\bar{q}_1}
+ \frac {m_{\bar{q}_2}}{m_{{q}_1}+m_{\bar{q}_2}}
\vec{r}_{q_2\bar{q}_2},
\end{equation}
\begin{equation}
\vec{r}_{\bar{q}_2} = \vec {R}_{q_1\bar{q}_2}
- \frac {m_{{q}_1}}{m_{{q}_1} + m_{\bar{q}_2}}
\vec{r}_{q_1\bar{q}_1}
- \frac {m_{{q}_1}}{m_{{q}_1}+m_{\bar{q}_2}}
\vec{r}_{q_2\bar{q}_2},
\end{equation}
which lead to
\begin{equation}
d\vec{r}_{q_1} d\vec{r}_{\bar{q}_1} d\vec{r}_{\bar{q}_2}=
d\vec{r}_{q_1\bar{q}_1} d\vec{r}_{q_2\bar{q}_2} d\vec{R}_{\rm total}.
\end{equation}
Using the independent position vectors and Eq. (23), we have for the left lower
diagram:
\begin{eqnarray}
& & <H \mid V_{{\rm r}q_2\bar{q}_1 q_1}\mid A,B>
=<q_1\bar{q}_2 \mid V_{{\rm r}q_2\bar{q}_1q_1} \mid q_1\bar{q}_1, q_2\bar{q}_2>
    \nonumber    \\
& = & \int d\vec{r}_{q_1} d\vec{r}_{\bar{q}_1} d\vec{r}_{\bar{q}_2}
\frac {e^{-i\vec{P}_{\rm f}\cdot\vec{R}_{\rm total}}}{\sqrt V}
\psi_{q_1\bar{q}_2}^+ (\vec{r}_{q_1\bar{q}_2})
    \nonumber    \\
& &
V_{{\rm r}q_2 \bar{q}_1 q_1}
\frac {e^{i\vec{P}_{q_1\bar{q}_1}\cdot\vec{R}_{q_1\bar{q}_1}}}{\sqrt V}
\psi_{q_1\bar{q}_1}(\vec{r}_{q_1\bar{q}_1})
\frac {e^{i\vec{P}_{q_2\bar{q}_2}\cdot\vec{R}_{q_2\bar{q}_2}}}{\sqrt V}
\psi_{q_2\bar{q}_2}(\vec{r}_{q_2\bar{q}_2})
    \nonumber    \\
& = & \int d\vec{r}_{q_1\bar{q}_1} d\vec{r}_{q_2\bar{q}_2} d\vec{R}_{\rm total}
V^{-\frac {3}{2}}e^{-i\vec{P}_{\rm f}\cdot\vec{R}_{\rm total}}
\psi_{q_1\bar{q}_2}^+ (\vec{r}_{q_1\bar{q}_2})
    \nonumber    \\
& &
V_{{\rm r}q_2 \bar{q}_1 q_1}\psi_{q_1\bar{q}_1} (\vec{r}_{q_1\bar{q}_1})
\psi_{q_2\bar{q}_2} (\vec{r}_{q_2\bar{q}_2})
e^{i\vec{P}_{\rm i}\cdot\vec{R}_{\rm total}
+i\vec{p}_{q_1\bar{q}_1,q_2\bar{q}_2}\cdot\vec{r}_{q_1\bar{q}_1,q_2\bar{q}_2}}
    \nonumber    \\
& = & (2\pi)^3 \delta^3 (\vec{P}_{\rm f} - \vec{P}_{\rm i})
\frac {{\cal M}_{{\rm r}q_2 \bar{q}_1 q_1}}
{V^{\frac {3}{2}}\sqrt{2E_A2E_B2E_H}},
\end{eqnarray}
with
\begin{eqnarray}
{\cal M}_{{\rm r}q_2 \bar{q}_1 q_1}
& = &
\sqrt {2E_A2E_B2E_H}
\int d\vec{r}_{q_1\bar{q}_1} d\vec{r}_{q_2\bar{q}_2}
\psi_{q_1\bar{q}_2}^+ (\vec {r}_{q_1\bar{q}_2})V_{{\rm r}q_2\bar{q}_1 q_1}
      \nonumber   \\
& &
\psi_{q_1\bar{q}_1} (\vec {r}_{q_1\bar {q}_1})
\psi_{q_2\bar{q}_2} (\vec {r}_{q_2\bar {q}_2})
e^{i\vec{p}_{q_1\bar{q}_1,q_2\bar{q}_2}
\cdot\vec{r}_{q_1\bar{q}_1,q_2\bar{q}_2}}.
\end{eqnarray}
For the right lower diagram we obtain
\begin{eqnarray}
<H \mid V_{{\rm r}q_2 \bar{q}_1 \bar{q}_2}\mid A,B> & = & 
<q_1\bar{q}_2 \mid V_{{\rm r}q_2\bar{q}_1\bar{q}_2} \mid 
q_1\bar{q}_1, q_2\bar{q}_2>
      \nonumber   \\
& = & (2\pi)^3 \delta^3 (\vec{P}_{\rm f} - \vec{P}_{\rm i})
\frac {{\cal M}_{{\rm r}q_2 \bar{q}_1 \bar{q}_2}}
{V^{\frac {3}{2}}\sqrt{2E_A2E_B2E_H}},
\end{eqnarray}
with
\begin{eqnarray}
{\cal M}_{{\rm r}q_2 \bar{q}_1 \bar{q}_2}
& = &
\sqrt {2E_A2E_B2E_H}
\int d\vec{r}_{q_1\bar{q}_1} d\vec{r}_{q_2\bar{q}_2}
\psi_{q_1\bar{q}_2}^+ (\vec {r}_{q_1\bar{q}_2})
V_{{\rm r}q_2 \bar{q}_1 \bar{q}_2}
      \nonumber   \\
& &
\psi_{q_1\bar{q}_1} (\vec {r}_{q_1\bar {q}_1})
\psi_{q_2\bar{q}_2} (\vec {r}_{q_2\bar {q}_2})
e^{i\vec{p}_{q_1\bar{q}_1,q_2\bar{q}_2}
\cdot\vec{r}_{q_1\bar{q}_1,q_2\bar{q}_2}}.
\end{eqnarray}

We take the Fourier transform of the meson wave functions and the transition
potentials:
\begin{equation}
\psi_{q_1\bar{q}_1}(\vec{r}_{q_1\bar{q}_1}) =
\int \frac {d^3p_{q_1\bar{q}_1}}{(2\pi)^3} \psi_{q_1\bar{q}_1}
(\vec {p}_{q_1\bar{q}_1})
e^{i\vec {p}_{q_1\bar{q}_1} \cdot \vec {r}_{q_1\bar{q}_1}},
\end{equation}
\begin{equation}
\psi_{q_2\bar{q}_2}(\vec{r}_{q_2\bar{q}_2}) =
\int \frac {d^3p_{q_2\bar{q}_2}}{(2\pi)^3} \psi_{q_2\bar{q}_2}
(\vec {p}_{q_2\bar{q}_2})
e^{i\vec {p}_{q_2\bar{q}_2} \cdot \vec {r}_{q_2\bar{q}_2}},
\end{equation}
\begin{equation}
\psi_{q_2\bar{q}_1}(\vec{r}_{q_2\bar{q}_1}) =
\int \frac {d^3p_{q_2\bar{q}_1}}{(2\pi)^3} \psi_{q_2\bar{q}_1}
(\vec {p}_{q_2\bar{q}_1})
e^{i\vec {p}_{q_2\bar{q}_1} \cdot \vec {r}_{q_2\bar{q}_1}},
\end{equation}
\begin{equation}
\psi_{q_1\bar{q}_2}(\vec{r}_{q_1\bar{q}_2}) =
\int \frac {d^3p_{q_1\bar{q}_2}}{(2\pi)^3} \psi_{q_1\bar{q}_2}
(\vec {p}_{q_1\bar{q}_2})
e^{i\vec {p}_{q_1\bar{q}_2} \cdot \vec {r}_{q_1\bar{q}_2}},
\end{equation}
\begin{equation}
V_{{\rm r}q_1\bar{q}_2\bar{q}_1}(\vec{r}_{\bar{q}_1}-\vec{r}_{q_1}) =
\int \frac {d^3k}{(2\pi)^3} V_{{\rm r}q_1\bar{q}_2\bar{q}_1} (\vec {k})
e^{i\vec {k} \cdot (\vec{r}_{\bar{q}_1}-\vec{r}_{q_1})},
\end{equation}
\begin{equation}
V_{{\rm r}q_1\bar{q}_2q_2}(\vec{r}_{q_2}-\vec{r}_{\bar{q}_2}) =
\int \frac {d^3k}{(2\pi)^3} V_{{\rm r}q_1\bar{q}_2q_2} (\vec {k})
e^{i\vec {k} \cdot (\vec{r}_{q_2}-\vec{r}_{\bar{q}_2})},
\end{equation}
\begin{equation}
V_{{\rm r}q_2\bar{q}_1q_1}(\vec{r}_{q_1}-\vec{r}_{\bar{q}_1}) =
\int \frac {d^3k}{(2\pi)^3} V_{{\rm r}q_2\bar{q}_1q_1} (\vec {k})
e^{i\vec {k} \cdot (\vec{r}_{q_1}-\vec{r}_{\bar{q}_1})},
\end{equation}
\begin{equation}
V_{{\rm r}q_2\bar{q}_1\bar{q}_2}(\vec{r}_{\bar{q}_2}-\vec{r}_{q_2}) =
\int \frac {d^3k}{(2\pi)^3} V_{{\rm r}q_2\bar{q}_1\bar{q}_2} (\vec {k})
e^{i\vec {k} \cdot (\vec{r}_{\bar{q}_2}-\vec{r}_{q_2})},
\end{equation}
where $\vec k$ is the gluon momentum. The normalizations are 
$\int d\vec{r}_{ab}\psi_{ab}^+(\vec{r}_{ab})\psi_{ab}(\vec{r}_{ab})=1$ and
$\int \frac{d^3p_{ab}}{(2\pi)^3}
\psi_{ab}^+(\vec{p}_{ab})\psi_{ab}(\vec{p}_{ab})=1$.
Substitute the Fourier transform in Eqs.
(17), (19), (25), and (27) to get
\begin{eqnarray}
{\cal M}_{{\rm r}q_1\bar{q}_2 \bar{q}_1}
& = &
\sqrt {2E_A2E_B2E_H}
\int \frac{d^3 p_{q_1\bar{q}_1}}{(2\pi)^3}\frac{d^3 p_{q_2\bar{q}_2}}{(2\pi)^3}
\psi_{q_2\bar{q}_1}^+ (\vec {p}_{q_2\bar{q}_1})V_{{\rm r}q_1\bar{q}_2 
\bar{q}_1} (\vec{k})
      \nonumber   \\
& &
\psi_{q_1\bar{q}_1} (\vec {p}_{q_1\bar {q}_1})
\psi_{q_2\bar{q}_2} (\vec {p}_{q_2\bar {q}_2}),
\end{eqnarray}
\begin{eqnarray}
{\cal M}_{{\rm r}q_1\bar{q}_2 q_2}
& = &
\sqrt {2E_A2E_B2E_H}
\int \frac{d^3 p_{q_1\bar{q}_1}}{(2\pi)^3}
\frac{d^3 p_{q_2\bar{q}_2}}{(2\pi)^3}
\psi_{q_2\bar{q}_1}^+ (\vec {p}_{q_2\bar{q}_1})V_{{\rm r}q_1\bar{q}_2 q_2}
(\vec{k})
    \nonumber   \\
& &
\psi_{q_1\bar{q}_1} (\vec {p}_{q_1\bar {q}_1})
\psi_{q_2\bar{q}_2} (\vec {p}_{q_2\bar {q}_2}),
\end{eqnarray}
\begin{eqnarray}
{\cal M}_{{\rm r}q_2 \bar{q}_1 q_1}
& = &
\sqrt {2E_A2E_B2E_H}
\int \frac{d^3 p_{q_1\bar{q}_1}}{(2\pi)^3}\frac{d^3 p_{q_2\bar{q}_2}}{(2\pi)^3}
\psi_{q_1\bar{q}_2}^+ (\vec {p}_{q_1\bar{q}_2})V_{{\rm r}q_2 \bar{q}_1 q_1}
(\vec{k})
    \nonumber   \\
& &
\psi_{q_1\bar{q}_1} (\vec {p}_{q_1\bar {q}_1})
\psi_{q_2\bar{q}_2} (\vec {p}_{q_2\bar {q}_2}),
\end{eqnarray}
\begin{eqnarray}
{\cal M}_{{\rm r}q_2 \bar{q}_1 \bar{q}_2}
& = &
\sqrt {2E_A2E_B2E_H}
\int \frac{d^3 p_{q_1\bar{q}_1}}{(2\pi)^3}\frac{d^3 p_{q_2\bar{q}_2}}{(2\pi)^3}
\psi_{q_1\bar{q}_2}^+ (\vec {p}_{q_1\bar{q}_2})V_{{\rm r} q_2\bar{q}_1 
\bar{q}_2} (\vec{k})
    \nonumber   \\
& &
\psi_{q_1\bar{q}_1} (\vec {p}_{q_1\bar {q}_1})
\psi_{q_2\bar{q}_2} (\vec {p}_{q_2\bar {q}_2}).
\end{eqnarray}

The unpolarized cross section for $A+B \to H$ is
\begin{eqnarray}
\sigma^{\rm unpol} & = &
\frac {\pi\delta(E_{\rm f}-E_{\rm i})}
{4\sqrt {(P_A \cdot P_B)^2 - m_A^2m_B^2}E_H}\frac {1}{(2J_A+1)(2J_B+1)}
      \nonumber  \\
& &
\sum\limits_{J_{Az}J_{Bz}J_{Hz}}
\mid {\cal M}_{{\rm r}q_1\bar{q}_2 \bar{q}_1}+{\cal M}_{{\rm r}q_1\bar{q}_2q_2}
+{\cal M}_{{\rm r}q_2\bar{q}_1 q_1}
+{{\cal M}_{{\rm r}q_2\bar{q}_1 \bar{q}_2}}\mid^2 .
\end{eqnarray}
In the center-of-mass frame of the two initial mesons $E_{\rm f}=m_H$ and 
$E_{\rm i}=\sqrt{s}$. Then, $\delta (E_{\rm f}-E_{\rm i}) 
=\delta (E_{\rm i}-E_{\rm f}) = \delta (\sqrt{s}
-m_H)$. The experimental cross sections of $\pi \pi \to \rho$ are large
but finite. Since $\delta (x)=\lim\limits_{h \to 0}\frac {1}{h\sqrt{\pi}}\exp
(-\frac {x^2}{h^2})$, we approximate $ \delta (\sqrt{s}-m_H)$ by
$\frac {1}{h\sqrt{\pi}}\exp (-\frac {(\sqrt{s}-m_H)^2}{h^2})$ with a small 
value of $h$ to mimic the experimental data. Substituting the relation,
\begin{equation}
(P_A \cdot P_B)^2 -m_A^2m_B^2 = \frac{1}{4} [s-(m_A+m_B)^2][s-(m_A-m_B)^2],
\end{equation}
in Eq. (40), we get the unpolarized cross section,
\begin{eqnarray}
\sigma^{\rm unpol} & = &
\frac {\sqrt{\pi}\exp (-\frac {(\sqrt{s}-m_H)^2}{h^2})}
{2h\sqrt {[s-(m_A+m_B)^2][s-(m_A-m_B)^2]}E_H}\frac {1}{(2J_A+1)(2J_B+1)}
      \nonumber  \\
& &
\sum\limits_{J_{Az}J_{Bz}J_{Hz}}
\mid {\cal M}_{{\rm r}q_1\bar{q}_2 \bar{q}_1}+{\cal M}_{{\rm r}q_1\bar{q}_2q_2}
+{\cal M}_{{\rm r}q_2 \bar{q}_1 q_1}
+{{\cal M}_{{\rm r}q_2 \bar{q}_1 \bar{q}_2}}\mid^2 ,
\end{eqnarray}
with $\sqrt{s}=m_H$. Let $I_A$ ($I_B$, $I_H$) be the isospin of meson $A$ 
($B$, $H$). The isospin-averaged unpolarized cross section for $A+B \to H$ is
\begin{equation}
\sigma^{\rm un}=\frac {2I_H+1}{(2I_A+1)(2I_B+1)}\sigma^{\rm unpol}.
\end{equation}

\vspace{0.5cm}
\leftline{\bf IV. TRANSITION POTENTIAL AND TRANSITION AMPLITUDE}
\vspace{0.5cm}

The reaction $A+B \to H$ involves quark-antiquark annihilation into a gluon
and subsequent gluon absorption by a quark or an antiquark. This process is
shown in Fig. 2 where the left diagram indicates 
$q^\prime (p_1) + q(p_2) + \bar{q}(-p_3) \to q^\prime (p_1^\prime)$ 
and the right diagram indicates
$\bar{q}^\prime (-p_1) + q(p_2) + \bar{q}(-p_3) \to 
\bar{q}^\prime (-p_1^\prime)$.
The left diagram has been used to study $p\bar p$ annihilation into two mesons
\cite{KW}. According to the Feynman rules in QCD \cite{Muta}, the amplitude 
for the left diagram in Fig. 2 is written as
\begin{equation}
{\cal M}_{{\rm r}q\bar{q}q^\prime}=\frac{{g_s}^2}{k^2}\bar{\psi}_{q'}
(\vec{p}^{~\prime}_1,s^{\prime}_{q'z})\gamma_\tau T^e\psi_{q'}
(\vec{p}_1,s_{q'z})
\bar{\psi}_{\bar q}(\vec{p}_3,s_{\bar{q}z})\gamma^\tau T^e \psi_q
(\vec{p}_2,s_{qz}) ,
\end{equation}
where $g_{\rm s}$ is the gauge coupling constant; the gluon has the 
four-momentum $k$, the color index $e$, and the space-time index $\tau$;
$T^e$ $(e=1,\cdot \cdot \cdot,8)$ are the $SU(3)$ color generators;
$\gamma^\tau$ are the Dirac matrices; repeated color and space-time
indices ($e$ and $\tau$) are summed. The quark spinors 
($\psi_{q'}(\vec{p}_1,s_{q'z})$, 
$\psi_{q'}(\vec{p}^{~\prime}_1,s^{\prime}_{q'z})$,
$\psi_q (\vec{p}_2,s_{qz})$) and the antiquark spinor
($\psi_{\bar q} (\vec{p}_3,s_{\bar{q}z})$) are given in a familiar way 
\cite{SXW,BD} by
\begin{equation}
\psi_q(\vec{p}_2,s_{qz})=\left(
\begin{array}{ccc} 
G_2(\vec {p}_2)\\{\frac{\vec\sigma\cdot\vec {p}_2}{2m_q}}G_2(\vec{p}_2)
\end{array}
\right)\chi_{s_{qz}},
\end{equation}
\begin{equation}
\psi_{\bar q}(\vec{p}_3,s_{\bar{q}z})=\left(
\begin{array}{ccc}
\frac{\vec\sigma\cdot\vec{p}_3}{2m_{\bar q}}G_3(\vec{p}_3)\\G_3(\vec{p}_3)
\end{array}
\right)\chi_{s_{\bar{q}z}},
\end{equation}
\begin{equation}
\psi_{q^{\prime}}(\vec{p}_1,s_{q'z})=\left(
\begin{array}{ccc}
G_1(\vec{p}_1)\\{\frac{\vec\sigma\cdot\vec{p}_1}{2m_{q'}}}G_1(\vec{p}_1)
\end{array}
\right)\chi_{s_{q'z}},
\end{equation}
\begin{equation}
\psi_{q'}(\vec{p}^{~\prime}_1,s^{\prime}_{q'z})=\left(
\begin{array}{ccc}
G^{\prime}_1(\vec{p}^{~\prime}_1)\\{\frac{\vec\sigma\cdot\vec{p}^{~\prime}_1}
{2m_{q'}}} G^{\prime}_1(\vec{p}^{~\prime}_1)
\end{array}
\right)\chi_{s^{\prime}_{q'z}},
\end{equation}
where $\vec \sigma$ are the Pauli matrices, 
$\chi_{s_{qz}}$, $\chi_{s_{\bar{q}z}}$, $\chi_{s_{q'z}}$, and 
$\chi_{s^{\prime}_{q'z}}$ are the spin wave functions with the magnetic
projection quantum numbers, $s_{qz}$, $s_{\bar{q}z}$, $s_{q'z}$, and 
$s^{\prime}_{q'z}$, of the quark or antiquark spin, respectively.
The amplitude for the right diagram in Fig. 2 is,
\begin{equation}
{\cal M}_{{\rm r}q\bar{q}\bar{q}^{\prime}}=-\frac{{g_s}^2}{k^2}
{\bar{\psi}_{\bar{q}^{\prime}}}(\vec{p}_1,{s_{\bar{q}'z}})\gamma_{\tau}
T^e\psi_{\bar{q}'}(\vec{p}^{~\prime}_1,s^{\prime}_{\bar{q}'z})
\bar{\psi}_{\bar q}(\vec{p}_3,s_{\bar{q}z})\gamma^{\tau}
T^e\psi_q(\vec{p}_2,s_{qz}) ,
\end{equation}
where the antiquark spinors, $\psi_{\bar{q}'}(\vec{p}_1,s_{\bar{q}'z})$ and 
$\psi_{\bar{q}'}(\vec{p}^{~\prime}_1,s^{\prime}_{\bar{q}'z})$, are given by
\begin{equation}
\psi_{\bar{q}'}(\vec{p}_1,s_{\bar{q}'z})=\left(
\begin{array}{ccc}
\frac{\vec\sigma\cdot\vec p_1}{2m_{\bar{q}'}}G_1(\vec {p}_1)\\G_1(\vec {p}_1)
\end{array}
\right)\chi_{s_{\bar{q}'z}},
\end{equation}
\begin{equation}
\psi_{\bar{q}'}(\vec{p}^{~\prime}_1,s^{\prime}_{\bar{q}'z})=\left(
\begin{array}{ccc}
\frac{\vec\sigma\cdot\vec{p}^{~\prime}_1}{2m_{\bar{q}'}}
G^{\prime}_1(\vec{p}^{~\prime}_1)\\
G^{\prime}_1(\vec {p}^{~\prime}_1)
\end{array}
\right)\chi_{s^{\prime}_{\bar{q}'z}},
\end{equation}
where $\chi_{s_{\bar{q}'z}}$ and $\chi_{s^{\prime}_{\bar{q}'z}}$ are 
the spin wave functions with the magnetic projection quantum numbers, 
$s_{\bar{q}'z}$ and $s^{\prime}_{\bar{q}'z}$, of the antiquark spin, 
respectively.
Keeping such terms to order of the inverse of the quark mass, we get
\begin{eqnarray}
{\cal M}_{{\rm r}q\bar{q}q'}&=&\frac{g_s^2}{k^2}\chi^+_{s^{\prime}_{q'z}}
\chi^+_{s_{\bar{q}z}}T^eT^eG^{\prime}_1(\vec {p}^{~\prime}_1)G_3(\vec{p}_3)
\nonumber\\
& &
[\frac{\vec\sigma(32)\cdot\vec k}{2m_q}-\frac{\vec\sigma(1)\cdot\vec\sigma(32)
\vec\sigma(1)\cdot\vec {p}_1+\vec\sigma(1)\cdot\vec{p}^{~\prime}_1\vec\sigma(1)
\cdot\vec\sigma(32)}{2m_{q'}}]
\nonumber\\
& &
G_1(\vec {p}_1)G_2(\vec {p}_2)\chi_{s_{q'z}}\chi_{s_{qz}} ,
\end{eqnarray}
\begin{eqnarray}
{\cal M}_{{\rm r}q\bar{q}\bar {q}'}&=&-\frac{g_s^2}{k^2}
\chi^+_{s_{\bar{q}^{\prime}z}}
\chi^+_{s_{\bar{q}z}}T^eT^eG_1(\vec {p}_1)G_3(\vec {p}_3)
\nonumber\\
& &
[\frac{\vec\sigma(32)\cdot\vec k}{2m_q}
-\frac{\vec\sigma(1)\cdot\vec{p}_1\vec\sigma(1)\cdot\vec\sigma(32)
+\vec\sigma(1)\cdot\vec\sigma(32)\vec\sigma(1)\cdot\vec {p}^{~\prime}_1}
{2m_{\bar{q}'}}]
\nonumber\\
& &
G^{\prime}_1(\vec {p}^{~\prime}_1)G_2(\vec {p}_2)\chi_{s^{\prime}_{\bar{q}'z}}
\chi_{s_{qz}} .
\end{eqnarray}
Since $T^eT^e=\frac{\vec{\lambda}(1)}{2}\cdot\frac{\vec{\lambda}(32)}{2}$ 
with $\vec \lambda$ being the Gell-Mann matrices, we
obtain the transition potential for $q^\prime (p_1) + q(p_2) + \bar{q}(-p_3)
\to {q}^\prime (p_1^\prime)$,
\begin{equation}
V_{{\rm r}q\bar{q}q'}(\vec{k}) =
\frac{\vec{\lambda}(1)}{2}\cdot\frac{\vec{\lambda}(32)}{2}
\frac{g_{\rm s}^2}{k^2}
\left(\frac{\vec{\sigma}(32)\cdot\vec{k}}{2m_q}-
\frac{\vec{\sigma}(1)\cdot\vec{\sigma}(32)\vec{\sigma}(1)\cdot\vec{p}_1+
\vec{\sigma}(1)\cdot\vec{p}_1^{~\prime} \vec{\sigma}(1)\cdot\vec{\sigma}(32)}
{2m_{q^\prime}}\right) ,
\end{equation}
and the transition potential for $\bar{q}^\prime (-p_1) + q(p_2) +
\bar{q}(-p_3) \to \bar{q}^\prime (-p_1^\prime)$,
\begin{equation}
V_{{\rm r}q\bar{q}\bar{q}'}(\vec{k}) = -
\frac{\vec{\lambda}(1)}{2}\cdot\frac{\vec{\lambda}(32)}{2}
\frac{g_{\rm s}^2}{k^2}
\left(\frac{\vec{\sigma}(32)\cdot\vec{k}}{2m_q}-
\frac{\vec{\sigma}(1)\cdot\vec{p}_1 \vec{\sigma}(1)\cdot\vec{\sigma}(32)
+\vec{\sigma}(1)\cdot\vec{\sigma}(32)\vec{\sigma}(1)\cdot\vec{p}_1^{~\prime} }
{2m_{\bar{q}^\prime}}\right) .
\end{equation}
In Eqs. (54) and (55), $\vec{\lambda}(32)$ ($\vec{\sigma}(32)$)
mean that they have matrix elements 
between the color (spin) wave functions of the initial antiquark
and the initial quark.
In Eq. (54), $\vec{\lambda}(1)$ ($\vec{\sigma}(1)$) mean that they 
have matrix elements between the color (spin) wave functions of the final 
quark and the initial quark.
In Eq. (55), $\vec{\lambda}(1)$ ($\vec{\sigma}(1)$) mean that they 
have matrix elements between the color (spin) wave functions of the initial 
antiquark and the final antiquark.

Starting from the quark and antiquark spinors and the gluon-quark vertices 
with the Dirac matrices, the amplitude for the left or right diagram in Fig. 2
is proportional to the product of the transition potential and 
the nonrelativistic 
quark and antiquark wave functions ($G_1(\vec{p}_1)$, $G_2(\vec{p}_2)$,
$G_3(\vec{p}_3)$, $G_1^\prime(\vec{p}_1^{~\prime})$) 
in Eqs. (52) and (53). Only in obtaining the transition 
potential we use the quark and antiquark spinors. Because the nonrelativistic
wave functions appear in Eqs. (52) and (53), we proceed to get the $S$-matrix
element in Sect. III. In the nonrelativistic framework the product of the
nonrelativistic quark and antiquark wave functions in a meson is rewritten 
as the product of the quark-antiquark relative-motion wave function and 
the center-of-mass motion wave function as seen in Eqs. (9)-(11). 
Normalization in the volume $V$ should be given to the wave functions of meson
$H$ in Eq. (10) or (11), meson $A$, and meson $B$ in Eq. (9).

The wave function of mesons $A$ and $B$ is
\begin{equation}
\psi_{AB} =\phi_{A\rm rel} \phi_{B\rm rel} \phi_{A\rm color} \phi_{B\rm color}
\chi_{S_A S_{Az}} \chi_{S_B S_{Bz}} \varphi_{AB\rm flavor},
\end{equation}
and the wave function of meson $H$ is
\begin{equation}
\psi_H =\phi_{H\rm rel} \phi_{H\rm color} \chi_{S_H S_{Hz}} \phi_{H\rm flavor},
\end{equation}
where $\psi_{AB}=\psi_{q_1\bar{q}_1}\psi_{q_2\bar{q}_2}$;
$\psi_H=\psi_{q_2\bar{q}_1}=\psi_{q_1\bar{q}_2}$; $\phi_{A\rm rel}$
($\phi_{B\rm rel}$, $\phi_{H\rm rel}$),
$\phi_{A\rm color}$ ($\phi_{B\rm color}$, $\phi_{H\rm color}$), 
and $\chi_{S_A S_{Az}}$ ($\chi_{S_B S_{Bz}}$, $\chi_{S_H S_{Hz}}$)
are the quark-antiquark relative-motion wave function, the color wave function,
and the spin wave function of meson $A$ ($B$, $H$), respectively;
$\phi_{H\rm flavor}$ is the flavor wave function of meson $H$. The flavor
wave function $\varphi_{AB\rm flavor}$ of mesons $A$ and $B$ possesses the same
isospin as meson $H$. 
The spin of meson $A$ ($B$, $H$) is $S_A$ ($S_B$, $S_H$) with its magnetic 
projection quantum number $S_{Az}$ ($S_{Bz}$, $S_{Hz}$).
The transition amplitudes include color, spin, and
flavor matrix elements. The color matrix element is $-\frac {4}{3\sqrt 3}$,
$\frac {4}{3\sqrt 3}$, $\frac {4}{3\sqrt 3}$, and $-\frac {4}{3\sqrt 3}$
for the left upper diagram, the right upper diagram, the left lower diagram,
and the right lower diagram, respectively. 
The spin matrix elements in ${\cal M}_{{\rm r}q_1\bar{q}_2\bar{q}_1}$ 
and ${\cal M}_{{\rm r}q_1\bar{q}_2q_2}$ are listed in Table 1.
The spin matrix elements in ${\cal M}_{{\rm r}\bar{q}_1q_2q_1}$ 
and ${\cal M}_{{\rm r}\bar{q}_1q_2\bar{q}_2}$ equal the ones in
${\cal M}_{{\rm r}q_1\bar{q}_2q_2}$ and
${\cal M}_{{\rm r}q_1\bar{q}_2\bar{q}_1}$, respectively.
The flavor matrix element of $\pi \pi \to \rho$ ($\pi K \to K^*$) is 1 (0)
for the two upper diagrams and -1 
(-$\frac {\sqrt 6}{2}$) for the two lower diagrams.

\vspace{0.5cm}
\leftline{\bf V. NUMERICAL CROSS SECTIONS AND DISCUSSIONS }
\vspace{0.5cm}

We consider the following inelastic meson-meson scattering processes that are
governed not only by quark interchange but also by quark-antiquark annihilation
and creation:
\begin{eqnarray}
I=1/2~ \pi K^* \to \rho K, I=1/2~ \pi K^* \to \rho K^*,
I=1/2~ \pi K \to \rho K^*, I=1/2~ \rho K \to \rho K^*.
\nonumber
\end{eqnarray}
The flavor matrix elements of the four channels are -1/2 in
${\cal M}_{\rm fi}^{\rm prior}$ and ${\cal M}_{\rm fi}^{\rm post}$, 0 in
${\cal M}_{{\rm a}q_1\bar{q}_2}$, and 3/2 in
${\cal M}_{{\rm a}\bar{q}_1q_2}$. According to Eq.~(7)
we calculate unpolarized cross sections at the six temperatures
$T/T_{\rm c} =0$, $0.65$, $0.75$, $0.85$, $0.9$, and $0.95$, where $T_{\rm c}$
is the critical temperature and equals 0.175 GeV. In Figs. 3-6
we plot the unpolarized cross sections for the four channels of the reactions.
A quark-gluon plasma is produced in high-energy Au-Au collisions at  
the Relativistic Heavy Ion Collider. The quark-gluon plasma is not a phase of 
chiral symmetry restoration, but is a deconfined phase. The Au-Au collisions
undergo the phase transition between confinement and deconfinement.  
The transition temperature $T_{\rm c}$ measured from Au-Au collisions is 
0.175 GeV \cite{GLMRX}. This is consistent with the prediction of the lattice 
gauge calculations in Ref. \cite{KLP}. 

Numerical cross sections plotted in Figs. 3-6 are not convenient for use 
in future. Hence, the numerical cross sections should be parametrized.
All the cross sections shown
in Figs. 3-6 approach zero at $\sqrt{s} \rightarrow \infty$. 
In Figs. 3 and 4 every curve exhibits a peak. We may use a function 
of the form $a_1 (\frac{\sqrt{s}-\sqrt{s_0}}{b_1})^{e_1}
\exp [e_1(1-\frac{\sqrt{s}-\sqrt{s_0}}{b_1})]$ to fit the numerical 
cross sections, i.e., the curves. $\sqrt{s_0}$ is the threshold energy 
and decreases with increasing temperature.
The parameters $a_1$ and $b_1$ equal the height of the peak and the separation
between the peak's location on the $\sqrt s$-axis and the threshold energy, 
respectively. The function has only 
one maximum and can well fit some curves with one peak, but is insufficient 
for fitting a curve with
two peaks in Fig. 6. To remedy this, we use a sum of two functions,
\begin{eqnarray}
\sigma^{\rm unpol}(\sqrt {s},T)
&=&a_1 \left( \frac {\sqrt {s} -\sqrt {s_0}} {b_1} \right)^{e_1}
\exp \left[ e_1 \left( 1-\frac {\sqrt {s} -\sqrt {s_0}} {b_1} \right) \right]
\nonumber \\
&&+ a_2 \left( \frac {\sqrt {s} -\sqrt {s_0}} {b_2} \right)^{e_2}
\exp \left[ e_2 \left( 1-\frac {\sqrt {s} -\sqrt {s_0}} {b_2} \right) \right],
\end{eqnarray}
to get a satifactory fit. No more functions are used because of the terrible 
computation time. The values of the parameters, $a_1$, $b_1$, $e_1$, $a_2$,
$b_2$, and $e_2$, are listed in Table 2. The six parameters are positive and 
depend on temperature.
In the table $d_0$ is the separation between the peak's location on
the $\sqrt s$-axis and the threshold energy, and $\sqrt{s_{\rm z}}$ is
the square root of the Mandelstam variable at which the cross section is
1/100 of the peak cross section. 
At $\sqrt {s} =\sqrt {s_0}$ the parametrization gives zero 
cross section what is a feature of endothermic reactions.
Now $a_1$ does not equal the peak's height, and
the sum of $a_1$ and $a_2$ is mainly determined by the peak cross section.
The peak cross sections of the 4 reactions have the same behavior: from
$T/T_{\rm c}=0$ to 0.95 each first decreases to a value and then increases 
from the value. Correspondingly, $a_1$, $a_2$, and $a_1+a_2$ decrease first
and then generally increase. The parameters 
$b_1$ and $b_2$ are related to $d_0$. 
If $b_1 < b_2$, $\frac{d\sigma^{\rm unpol}}{d\sqrt s} > 0$ at 
$\sqrt{s} < b_1$ and $\frac{d\sigma^{\rm unpol}}{d\sqrt s} < 0$ at 
$\sqrt{s} > b_2$. If $b_1 > b_2$,
$\frac{d\sigma^{\rm unpol}}{d\sqrt s} > 0$ at $\sqrt s < b_2$ and 
$\frac{d\sigma^{\rm unpol}}{d\sqrt s} < 0$ at $\sqrt s > b_1$. 

Space and time are discretized in lattice QCD. Denote the temporal and 
spatial lattice spacings by $a_\tau$ and $a_\sigma$, respectively. The temporal
extent $N_\tau$ satisfies $N_\tau a_\tau = \beta \equiv \rm 1/T$. Let
$x=(\vec{x},x_4)$ label the space-time lattice sites.
With the link variables $U_\mu (\vec{x},x_4)$ \cite{Rothe} the Polyakov loop at
the spatial lattice site $\vec x$ is defined as 
\begin{eqnarray}
L (\vec{x})= Tr\displaystyle\prod_{x_4 = 1}^{N_\tau} U_4 (\vec{x},x_4).
\end{eqnarray}
The Polyakov loop correlation function is given by \cite{KLP}
\begin{eqnarray}
<L(\vec{x}_a)L^+(\vec{x}_b)> =\frac{\displaystyle\int\displaystyle\prod_{x\mu} 
dU_\mu (x) e^{-\beta S_G}L(\vec{x}_a)L^+(\vec{x}_b)
\displaystyle\prod_{q = 1}^{n_f}(\int
\displaystyle\prod_{x} d\bar{\chi}_x d\chi_x e^{-S_F})^{1/4}}
{\displaystyle\int\displaystyle\prod_{x\mu} dU_\mu (x) e^{-\beta S_G}
\displaystyle\prod_{q = 1}^{n_f}(\int
\displaystyle\prod_{x} d\bar{\chi}_x d\chi_x e^{-S_F})^{1/4}} ,
\end{eqnarray}
where $\chi_x$ is the transformed quark field \cite{Rothe}, $S_G$ is the gauge 
action \cite{KLP}, $S_F$ is the staggered fermion action \cite{KLP}, and
$n_f$ is the number of dynamical-quark flavors. The  
temperature dependence of the Polyakov loop
correlation function comes from the following two aspects. 
The first one is that $\beta$ explicitly gives rise to temperature 
dependence through $a_\tau$, $e^{-\beta S_G}$, and the two Polyakov loops.
The gluon field is subject to the periodic boundary condition that
its values at $x_4 = 0$ and $x_4 = N_\tau$ are equal. The quark
field satisfies the antiperiodic boundary condition that its values 
at $x_4 = 0$ and $x_4 = N_\tau$ differ only in sign. The second
is that the two
boundary conditions give rise to temperature dependence of the gluon field, 
the quark field, the link variable, and the integration measure.

Place a heavy quark at $\vec{x}_a$ and a heavy antiquark at $\vec{x}_b$.
The Polyakov loop correlation function is related to the free energy $F(T,r)$
of the heavy quark-antiquark pair by
\begin{equation}
-T\ln <L(\vec{x}_a)L^+(\vec{x}_b)>=F(T,r)+C' ,
\end{equation}
where $r=a_\sigma\mid \vec{x}_a - \vec{x}_b \mid$, and $C'$ is a normalization 
constant. In hadronic matter, i.e., $T<T_{\rm c}$ the quark-antiquark free
energy can be taken as the quark-antiquark potential 
\cite{RHS,BKR14,BKR16,BR,LMSK,SX2}. The lattice gauge calculations in Ref.
\cite{KLP} thus provide the numerical quark-antiquark potential at $r>0.3$ fm
at $T/T_{\rm c}>0.55$. The potential is temperature-dependent and 
spin-independent. At long distances the potential becomes a 
distance-independent value. With increasing temperature the value decreases.

The potential between constituents $a$ and $b$ in coordinate space is
\begin{eqnarray}
V_{ab}(\vec {r}) & = &
- \frac {\vec{\lambda}_a}{2} \cdot \frac {\vec{\lambda}_b}{2}
\frac {3}{4} D \left[ 1.3- \left( \frac {T}{T_{\rm c}} \right)^4 \right]
\tanh (Ar) + \frac {\vec{\lambda}_a}{2} \cdot \frac {\vec{\lambda}_b}{2}
\frac {6\pi}{25} \frac {v(\lambda r)}{r} \exp (-Er)
\nonumber  \\
& & -\frac {\vec{\lambda}_a}{2} \cdot \frac {\vec{\lambda}_b}{2}
\frac {16\pi^2}{25}\frac{d^3}{\pi^{3/2}}\exp(-d^2r^2) \frac {\vec {s}_a \cdot 
\vec {s} _b} {m_am_b}
+\frac {\vec{\lambda}_a}{2} \cdot \frac {\vec{\lambda}_b}{2}\frac {4\pi}{25}
\frac {1} {r}
\frac {d^2v(\lambda r)}{dr^2} \frac {\vec {s}_a \cdot \vec {s}_b}{m_am_b} ,
\end{eqnarray}
where $D=0.7$ GeV, $E=0.6$ GeV, $A=1.5[0.75+0.25 (T/{T_{\rm c}})^{10}]^6$ GeV, 
$\lambda=\sqrt{25/16\pi^2 \alpha'}$ with $\alpha'=1.04$ GeV$^{-2}$, 
$\lambda_a$ are the Gell-Mann matrices for the color generators of constituent
$a$, $\vec {s}_a$ is the spin of constituent $a$, the function $v$ is given by
Buchm\"uller and Tye in Ref. \cite{BT}, and the quantity $d$ given in Ref. 
\cite{SXW} is related to the constituent masses. The temperature dependence
of the potential is shown by the first term. The sum of the first and second
terms fits the lattice QCD results at intermediate and long distances at
$T/T_{\rm c}>0.55$ \cite{KLP}. Solving the Schr\"odinger equation with the
quark potential, we obtain the temperature dependence of meson masses and 
quark-antiquark relative-motion wave functions. The temperature-dependent
masses lead to temperature dependence of the threshold energy which is the
sum of the masses of the two final mesons for the 2-to-2 reactions or the 
mass of the final meson for the 2-to-1 reactions. The temperature dependence
of the potential, the meson masses, and the mesonic quark-antiquark wave
functions bring about the temperature dependence of the unpolarized cross
sections for the 2-to-2 reactions via the transition amplitudes in Eqs. 
(1)-(4),  while the masses and wave functions also do this for the 2-to-1 
reactions
via the transition amplitudes in Eqs. (36)-(39). We note that the transition
potentials in Eqs. (54)-(55) are temperature-independent.

The four reactions are endothermic. The cross section for  
$\pi K^* \to \rho K$ for $I=1/2$ at a given temperature 
increases very rapidly from threshold, reaches a maximum, and 
then decreases rapidly. However, the cross section for $\pi K \to \rho K^*$ for
$I=1/2$ at $T/T_{\rm c}=0$, 0.65, 0.75, or 0.85 decreases slowly from 
its maximum; the cross section for
$\rho K \to \rho K^*$ for $I=1/2$ at $T/T_{\rm c}=0.85$ even decreases first
and then increases. The cross section for $\rho K \to \rho K^*$ for $I=1/2$
at $T/T_{\rm c}=0.9$ or 0.95 has a narrow peak near threshold and a wide peak
around $\sqrt {s}=2.5$ GeV. 

The reaction $\pi K^* \to \rho K$ for 
$I=3/2$ is governed by quark interchange only \cite{SX1}.
The flavor matrix element of $\pi K^* \to \rho K$ for 
$I=3/2$ is 1, while the one of $\pi K^* \to \rho K$ for 
$I=1/2$ due to quark interchange
is -1/2. The peak cross section of $\pi K^* \to \rho K$ for $I=3/2$ at 
$T/T_{\rm c}=0.95$ is roughly four times the one of $\pi K^* \to \rho K$ for 
$I=1/2$ at the same temperature. Therefore, near the critical temperature
quark interchange dominates the reaction $\pi K^* \to \rho K$ for $I=1/2$.
This conclusion can also be drawn in the
cases of $\pi K^* \to \rho K^*$, $\pi K \to \rho K^*$, and 
$\rho K \to \rho K^*$.

The Buchm\"uller-Tye potential arises from one-gluon exchange plus 
perturbative one- and two-loop corrections \cite{BT}, and provides 
$g_{\rm s}=\frac{2\sqrt{6}\pi}{5}$ for one-gluon exchange.
Set $h=1/(53\sqrt{\pi})$ ${\rm fm}^{-1}$. According to Eq. (43) we calculate 
isospin-averaged unpolarized  cross sections for $\pi \pi \to \rho$ and
$\pi K \to K^\ast$ at various temperatures. Results are listed in Table
3. $\sigma^{\rm unpol}$ in Eq. (58) decreases when $\sqrt{s}-\sqrt{s_0}$
increases from the larger one of $b_1$ and $b_2$. It is shown in Table 3 that
the cross sections decrease when temperature increases. We thus use the
right-hand side of Eq. (58) to fit the numerical cross sections in Table 3 by
replacing $\sqrt{s}-\sqrt{s_0}$ with $T/T_{\rm c}-0.42$ for $\pi \pi \to \rho$
or $T/T_{\rm c}-0.5$ for $\pi K \to K^*$. In the temperature region 
$0.6 T_{\rm c} < T < T_{\rm c}$ hadronic matter exists.
The isospin-averaged unpolarized cross section for
$\pi \pi \to \rho$ for $0.6 T_{\rm c} \leq T < T_{\rm c}$ is parametrized as
\begin{eqnarray}
\sigma^{\rm un}(T)
&=& 54.1 \left( \frac {T/T_{\rm c} -0.42} {0.18} \right)^{3.3}
\exp \left[ 3.3 \left( 1-\frac {T/T_{\rm c} -0.42} {0.18} \right) \right]
\nonumber \\
&&+ 5.32 \left( \frac {T/T_{\rm c} -0.42} {0.5} \right)^{84}
\exp \left[ 84 \left( 1-\frac {T/T_{\rm c} -0.42} {0.5} \right) \right],
\end{eqnarray}
and for $\pi K \to K^*$,
\begin{eqnarray}
\sigma^{\rm un}(T)
&=& 42.2 \left( \frac {T/T_{\rm c} -0.5} {0.115} \right)^{1.83}
\exp \left[ 1.83 \left( 1-\frac {T/T_{\rm c} -0.5} {0.115} \right) \right]
\nonumber \\
&&+ 4.95 \left( \frac {T/T_{\rm c} -0.5} {0.407} \right)^{50}
\exp \left[ 50 \left( 1-\frac {T/T_{\rm c} -0.5} {0.407} \right) \right] .
\end{eqnarray}
At $T=0$ the cross sections for $\pi \pi \to \rho$ and
$\pi K \to K^\ast$ are 80.07 mb and 60.5 mb in comparison with the measured
value 80 mb \cite{FMMR} and the estimate 60 mb formed in Ref. \cite{Ko},
respectively.
The value 80.07 mb (60.5 mb) is much larger than 0.64 mb (0.27 mb) which 
is the maximum of
the isospin-averaged unpolarized cross section for $\pi \pi \to \rho \rho$
($\pi K \to \rho K^*$) at $T=0$ \cite{SX1,SXW}. With increasing temperature
the cross section for either reaction
shown in Table 3 decreases. As the temperature increases from
zero, the long-distance part of the quark potential $V_{ab}$
gradually becomes a distance-independent and temperature-dependent 
quantity. As the temperature increases from $0.6T_{\rm c}$ to $T_{\rm c}$, 
the quantity decreases and confinement becomes weaker and weaker. The weakening
confinement with increasing temperature makes combining the final quark and
the final antiquark into a meson more difficult, and thus reduces the cross
section. At $T/T_{\rm c}=0.95$ the cross
section for $\pi \pi \to \rho$ ($\pi K \to K^\ast$) is still much larger than 
the maximum of
the isospin-averaged unpolarized cross section for $\pi \pi \to \rho \rho$ 
($\pi K \to \rho K^\ast$), which is 0.11 mb (0.78 mb) at the same temperature
\cite{SX1,SXW}.

\vspace{0.5cm}
\leftline{\bf VI. SUMMARY }
\vspace{0.5cm}

We have provided cross section formulas for the reactions that are governed
by quark interchange as well as quark-antiquark annihilation and creation, and
have obtained the temperature dependence of the unpolarized cross sections for
$\pi K^* \to \rho K$ for $I=1/2$, $\pi K^* \to \rho K^\ast$ for $I=1/2$,
$\pi K \to \rho K^*$ for $I=1/2$, and $\rho K \to \rho K^*$ for $I=1/2$.
Near threshold quark interchange dominates the four channels near
the critical temperature; in the other energy region the quark-antiquark
annihilation and creation may dominate the four channels. 
The numerical cross sections are parametrized for 
future use in the evolution of hadronic matter.

We have proposed a model to study 2-to-1 meson-meson scattering.
The isospin-averaged unpolarized cross section for the scattering has been
derived, and the cross section formulas have been applied to study
$\pi \pi \to \rho$ and $\pi K \to K^\ast$. The reactions contain the process
where a quark-antiquark pair annihilates into a gluon and subsequently
the gluon is absorbed by a quark or an antiquark. The transition potential
of the process is derived from the Feynman rules
in perturbative QCD. The transition amplitudes
corresponding to the four Feynman diagrams are calculated with the
transition potential. The isospin-averaged
unpolarized cross sections for the two reactions decrease due to weakening
confinement with increasing 
temperature. The numerical cross sections are parametrized.

\vspace{0.5cm}
\leftline{\bf ACKNOWLEDGEMENTS}
\vspace{0.5cm}

This work was supported by the National Natural Science Foundation of China
under Grant No. 11175111.

\newpage
\begin{figure}[htbp]
  \centering
    \includegraphics[scale=0.84]{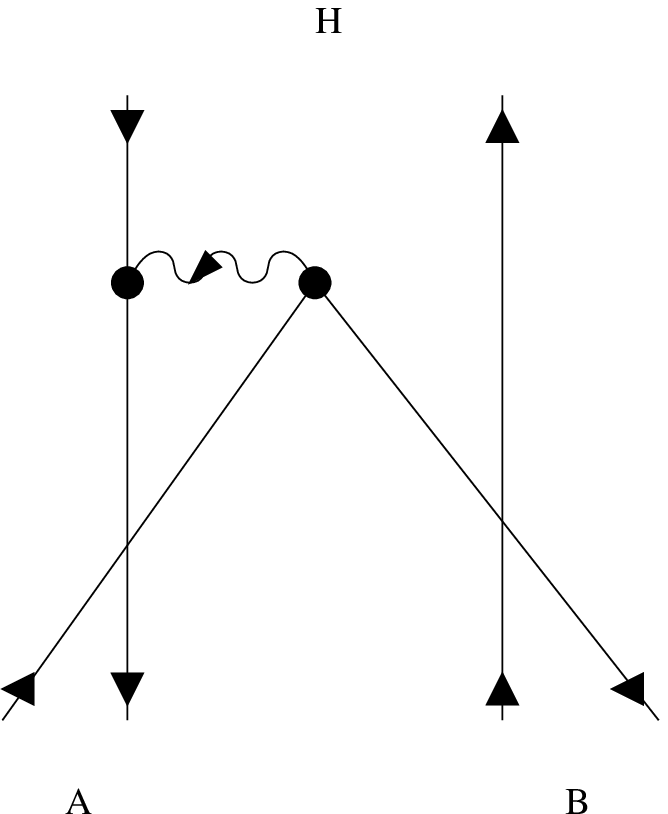}
      \hspace{3cm}
    \includegraphics[scale=0.84]{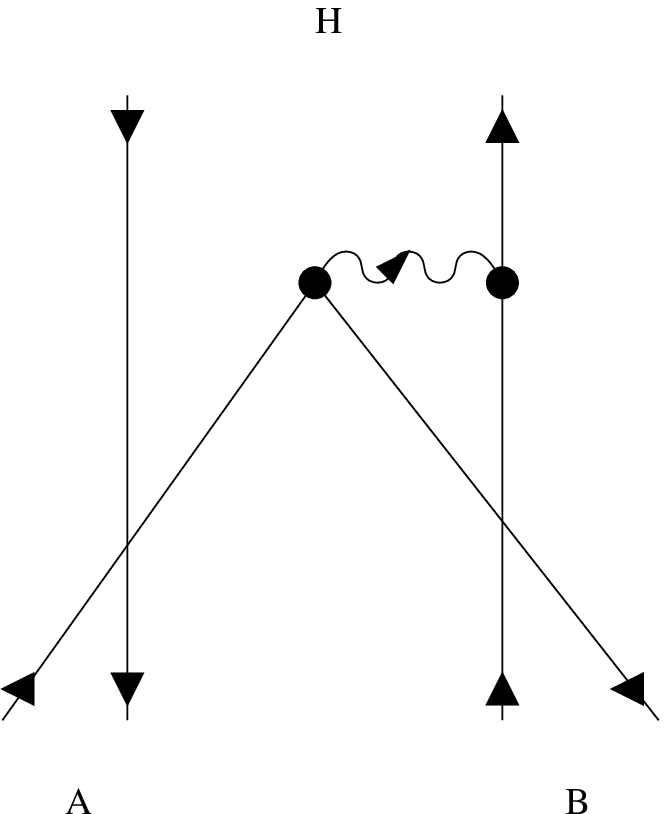}
      \vskip 72pt
    \includegraphics[scale=0.84]{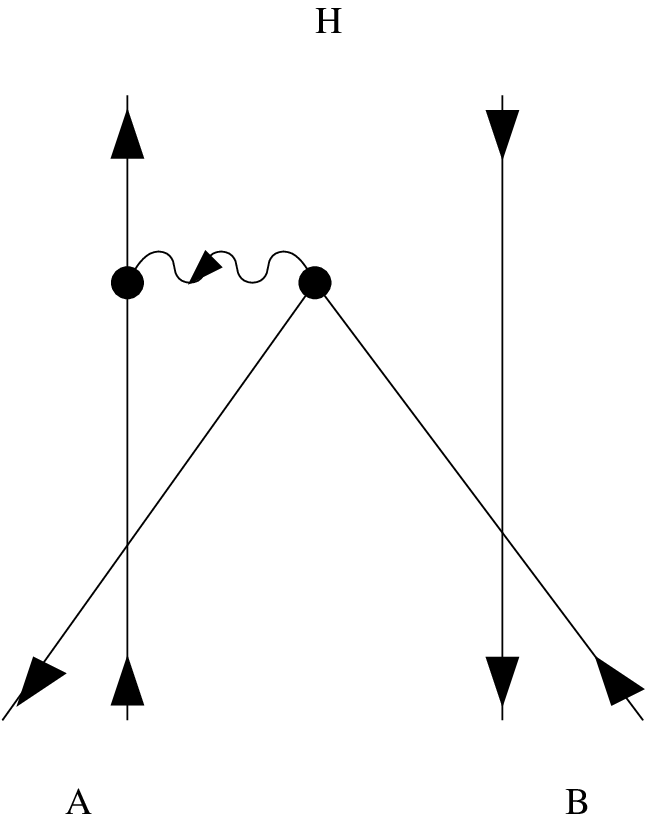}
      \hspace{3cm}
    \includegraphics[scale=0.84]{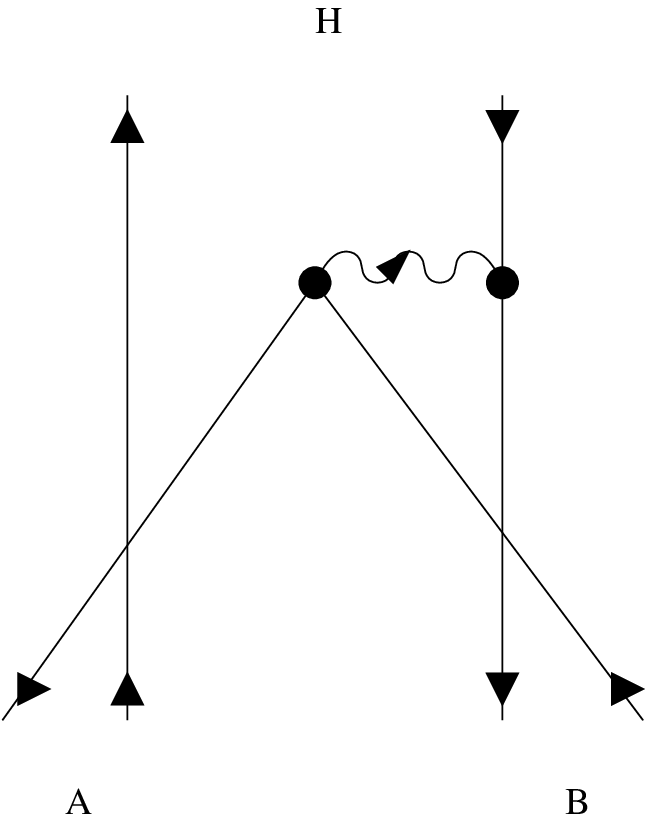}
\caption{Reaction $A+B \to H$. Solid lines with up (down) triangles represent
quarks (antiquarks). Wavy lines represent gluons.}
\label{fig1}
\end{figure}

\newpage
\begin{figure}[htbp]
  \centering
    \includegraphics[scale=0.84]{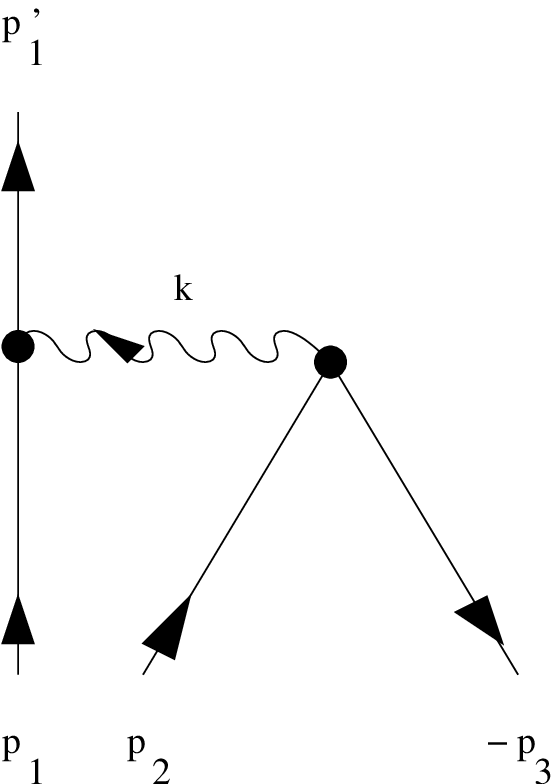}
      \hspace{2cm}
    \includegraphics[scale=0.84]{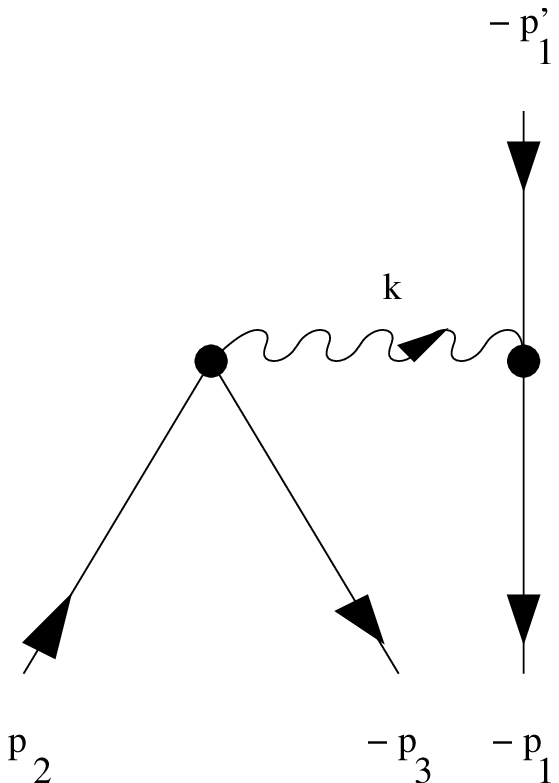}
\caption{Left diagram with $q'(p_1) + q(p_2) + \bar{q}(-p_3) \to 
q'(p_1^\prime)$ and right diagram with $\bar{q}'(-p_1) + q(p_2) + 
\bar {q}(-p_3) \to \bar{q}'(-p_1^\prime)$.}
\label{fig2}
\end{figure}

\newpage
\begin{figure}[htbp]
  \centering
    \includegraphics[scale=0.6]{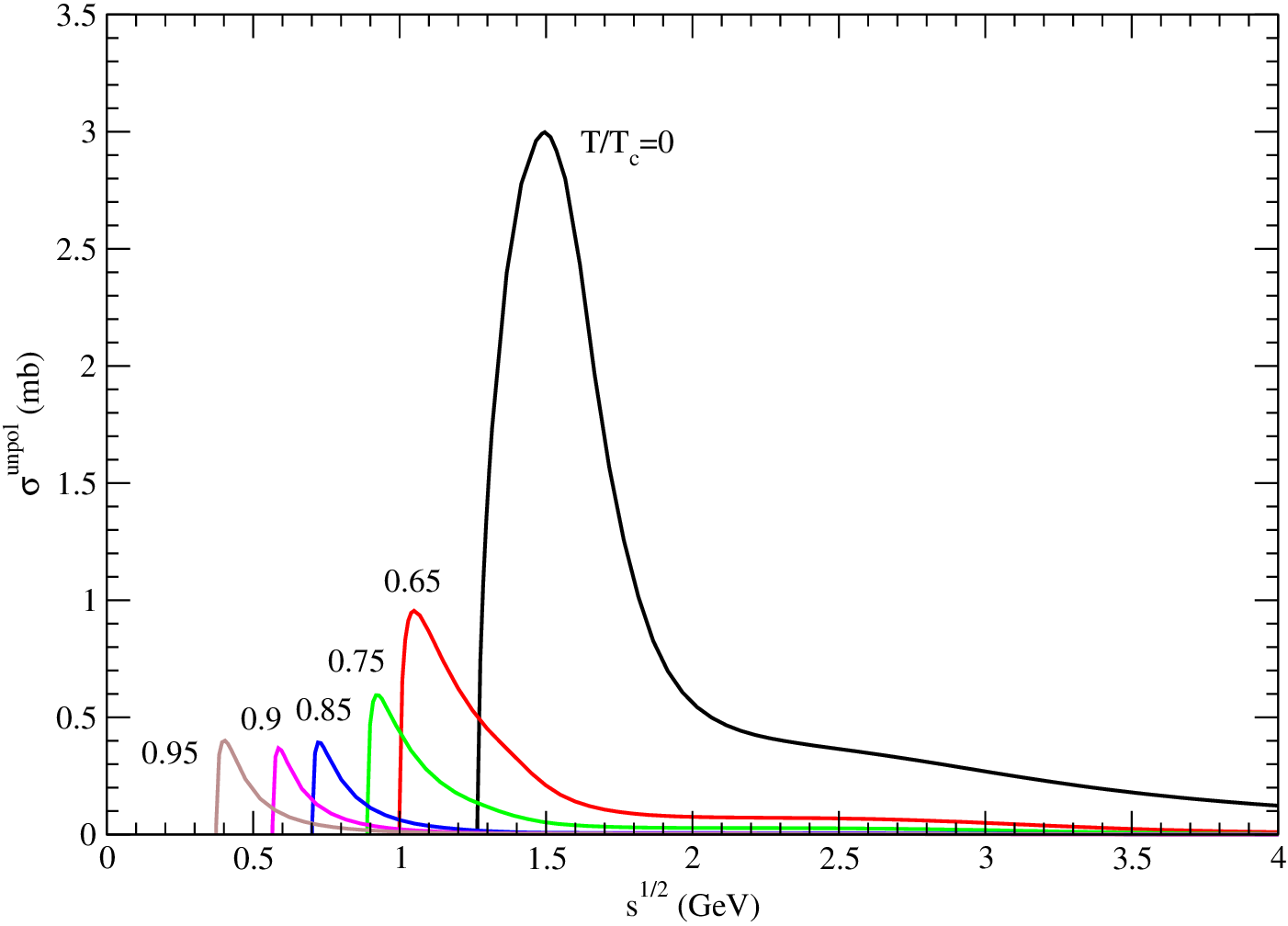}
\caption{Cross sections for $\pi K^* \to \rho K$ for $I=1/2$
at various temperatures.}
\label{fig3}
\end{figure}

\newpage
\begin{figure}[htbp]
  \centering
    \includegraphics[scale=0.65]{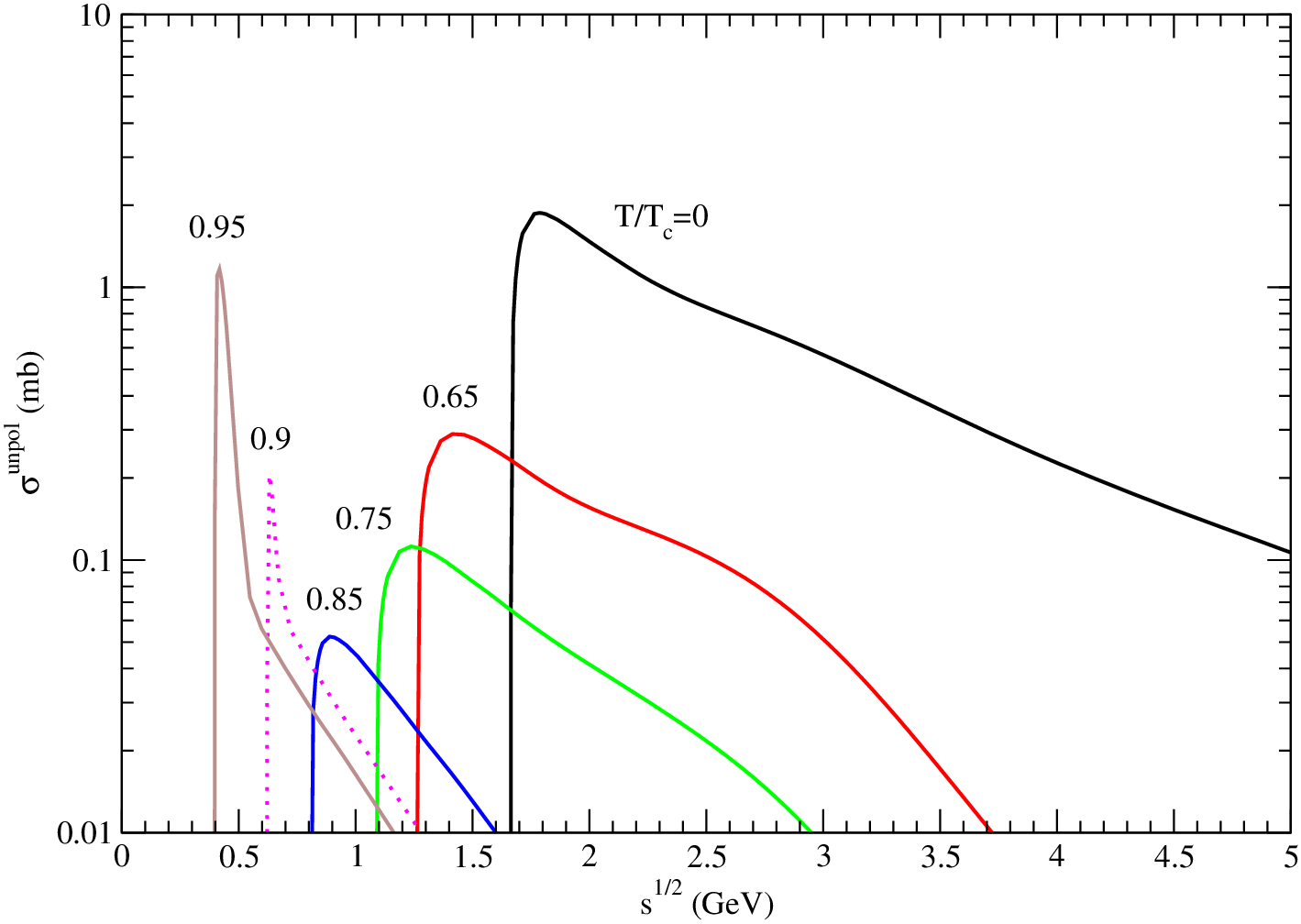}
\caption{Cross sections for $\pi K^* \to \rho K^*$ for $I=1/2$
at various temperatures.}
\label{fig4}
\end{figure}

\newpage
\begin{figure}[htbp]
  \centering
    \includegraphics[scale=0.65]{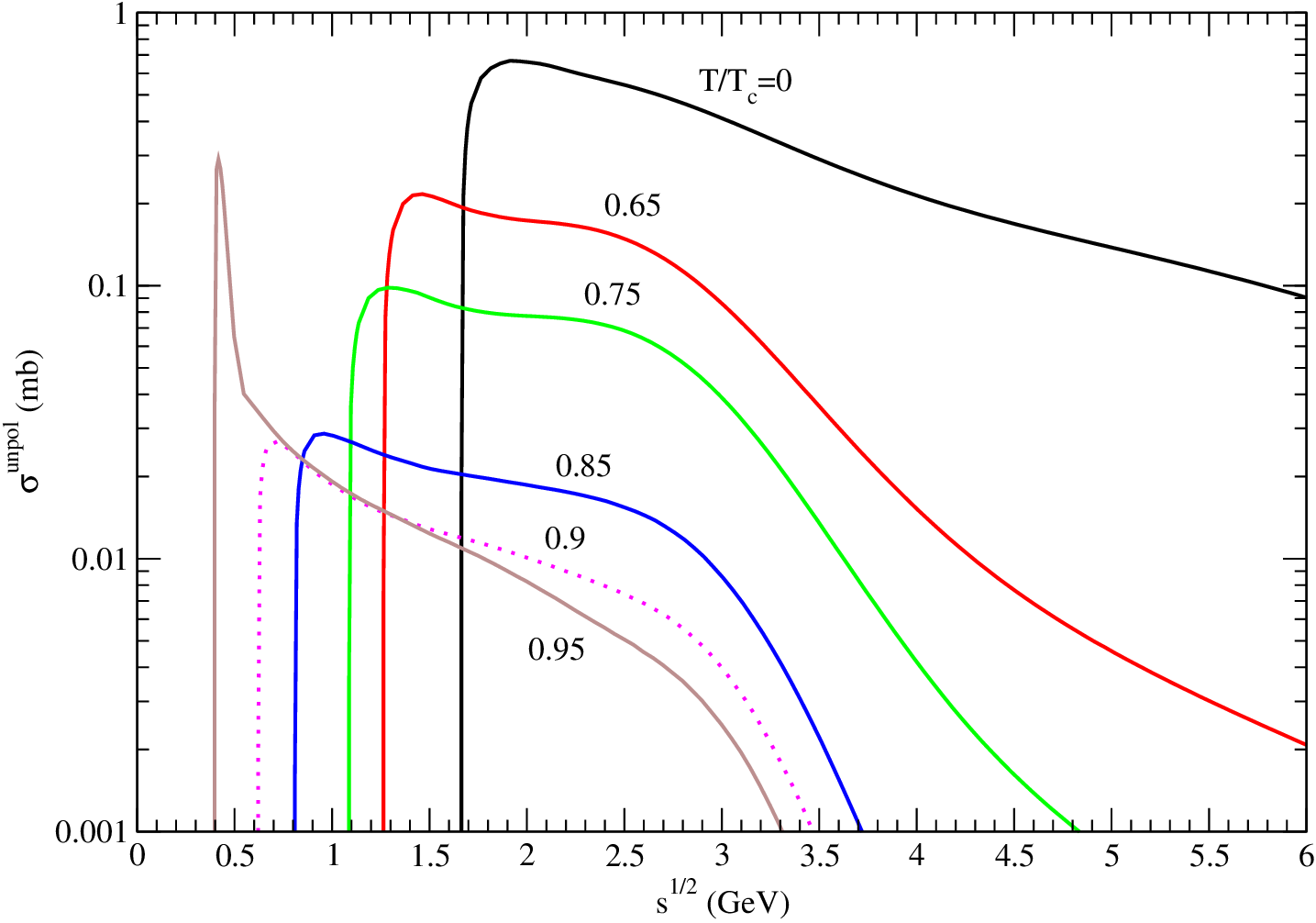}
\caption{Cross sections for $\pi K \to \rho K^*$ for $I=1/2$
at various temperatures.}
\label{fig5}
\end{figure}

\newpage
\begin{figure}[htbp]
  \centering
    \includegraphics[scale=0.65]{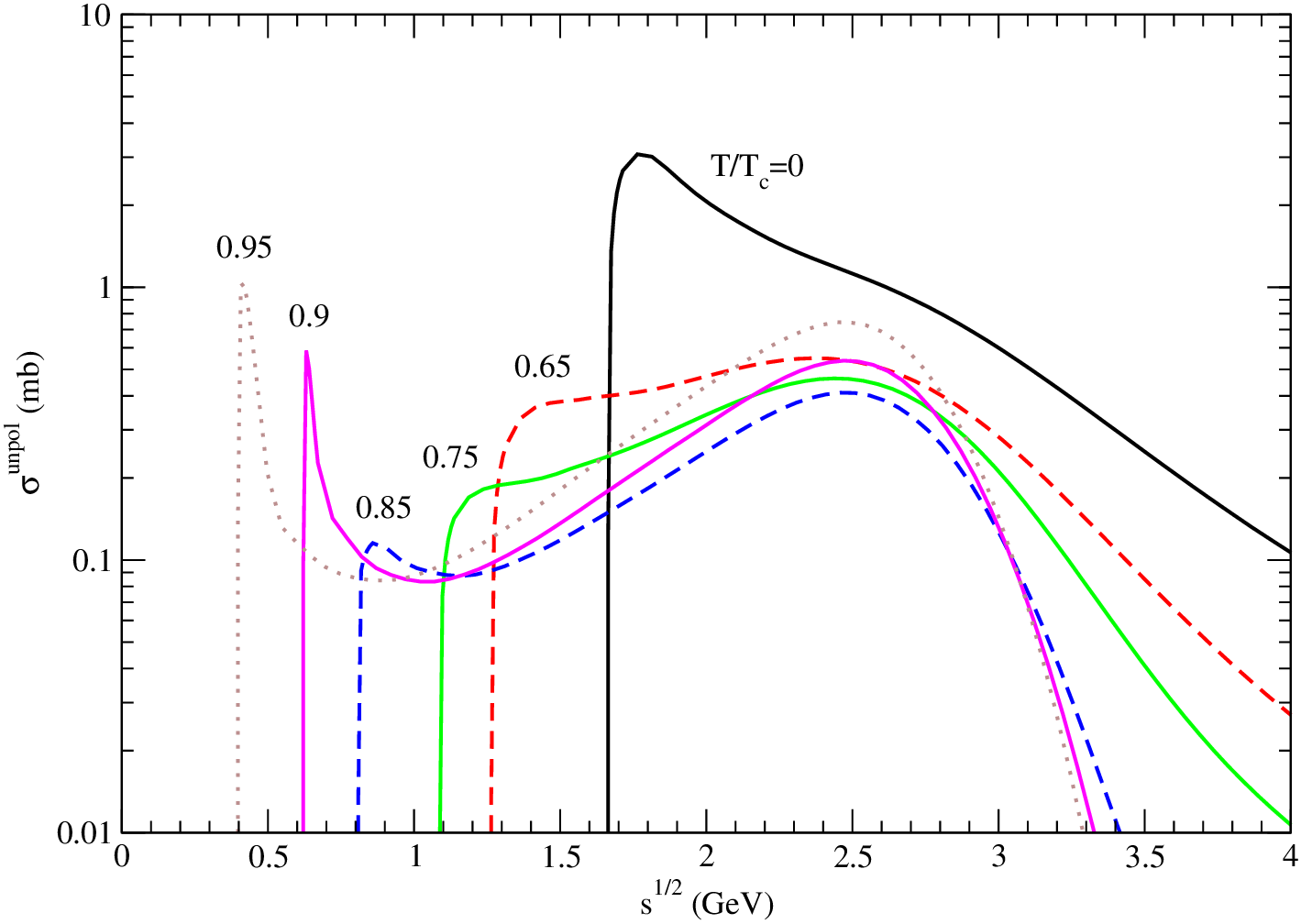}
\caption{Cross sections for $\rho K \to \rho K^*$ for $I=1/2$
at various temperatures.}
\label{fig6}
\end{figure}

\newpage
\begin{table}
\caption{\label{table1} Spin matrix elements in
${\cal M}_{{\rm r}q_1\bar{q}_2\bar{q}_1}$ and 
${\cal M}_{{\rm r}q_1\bar{q}_2q_2}$.
The initial spin state is
$\phi_{\rm iss}=\chi_{S_A S_{Az}} \chi_{S_B S_{Bz}}$, and
the final spin state $\phi_{\rm fss}=\chi_{S_H S_{Hz}}$.
The second to fourth columns correspond to 
${\cal M}_{{\rm r}q_1\bar{q}_2\bar{q}_1}$,
and the fifth to seventh columns ${\cal M}_{{\rm r}q_1\bar{q}_2q_2}$.}
\begin{tabular}{ccccccc}
\hline
$S_{Az}$ & 0 & 0  & 0  & 0 & 0  & 0 \\
$S_{Bz}$ & 0 & 0  & 0  & 0 & 0  & 0 \\
$S_{Hz}$ & -1 & 0 & 1  & -1 & 0 & 1 \\
\hline
$\phi_{\rm fss}^+ \phi_{\rm iss}$ & $-\frac {1}{2}$ & 0 & $-\frac {1}{2}$ 
& $-\frac {1}{2}$ & 0 & $-\frac {1}{2}$ \\
$\phi_{\rm fss}^+ \sigma_1(32) \phi_{\rm iss}$ & 0 & $\frac {1}{\sqrt 2}$ & 0 
& 0 & $\frac {1}{\sqrt 2}$ & 0 \\
$\phi_{\rm fss}^+ \sigma_2(32) \phi_{\rm iss}$ & 0 & 0 & 0 & 0 & 0 & 0 \\
$\phi_{\rm fss}^+ \sigma_3(32) \phi_{\rm iss}$ & $-\frac {1}{2}$ & 0 
& $\frac {1}{2}$ & $-\frac {1}{2}$ & 0 & $\frac {1}{2}$ \\
$\phi_{\rm fss}^+ \sigma_1(1) \phi_{\rm iss}$ & 0 & $-\frac {1}{\sqrt 2}$ & 0 
& 0 & $-\frac {1}{\sqrt 2}$ & 0 \\
$\phi_{\rm fss}^+ \sigma_2(1) \phi_{\rm iss}$ & 0 & 0 & 0 & 0 & 0 & 0 \\
$\phi_{\rm fss}^+ \sigma_3(1) \phi_{\rm iss}$ & $\frac {1}{2}$ & 0 
& $-\frac {1}{2}$ & $\frac {1}{2}$ & 0 & $-\frac {1}{2}$\\
\hline
$\phi_{\rm fss}^+ \sigma_1(32) \sigma_1(1) \phi_{\rm iss}$
& $\frac {1}{2}$ & 0 & $\frac {1}{2}$ & $\frac {1}{2}$ & 0 & $\frac {1}{2}$ \\
$\phi_{\rm fss}^+ \sigma_1(32) \sigma_2(1) \phi_{\rm iss}$
& -$\frac {1}{2}i$ & 0 & $\frac {1}{2}i$ & $\frac {1}{2}i$ & 0 
& -$\frac {1}{2}i$ \\
$\phi_{\rm fss}^+ \sigma_1(32) \sigma_3(1) \phi_{\rm iss}$
& 0 & 0 & 0 & 0 & 0 & 0 \\
$\phi_{\rm fss}^+ \sigma_2(32) \sigma_1(1) \phi_{\rm iss}$
& -$\frac {1}{2}i$ & 0 & $\frac {1}{2}i$ & $\frac {1}{2}i$ & 0 
& -$\frac {1}{2}i$ \\
$\phi_{\rm fss}^+ \sigma_2(32) \sigma_2(1) \phi_{\rm iss}$
& -$\frac {1}{2}$ & 0 & -$\frac {1}{2}$ & -$\frac {1}{2}$ & 0 
& -$\frac {1}{2}$ \\
$\phi_{\rm fss}^+ \sigma_2(32) \sigma_3(1) \phi_{\rm iss}$
& 0 & -$\frac {1}{\sqrt 2}i$ & 0 & 0 & $\frac {1}{\sqrt 2}i$ & 0 \\
$\phi_{\rm fss}^+ \sigma_3(32) \sigma_1(1) \phi_{\rm iss}$
& 0 & 0 & 0 & 0 & 0 & 0 \\
$\phi_{\rm fss}^+ \sigma_3(32) \sigma_2(1) \phi_{\rm iss}$
& 0 & -$\frac {1}{\sqrt 2}i$ & 0 & 0 & $\frac {1}{\sqrt 2}i$ & 0 \\
$\phi_{\rm fss}^+ \sigma_3(32) \sigma_3(1) \phi_{\rm iss}$
& $\frac {1}{2}$ & 0 & $\frac {1}{2}$ & $\frac {1}{2}$ & 0 & $\frac {1}{2}$ \\
\hline
\end{tabular}
\end{table}

\newpage
\begin{table*}[htbp]
\caption{\label{table2}Values of the parameters. $a_1$ and $a_2$ are
in units of millibarns; $b_1$, $b_2$, $d_0$, and $\sqrt{s_{\rm z}}$ are
in units of GeV; $e_1$ and $e_2$ are dimensionless.}
\tabcolsep=5pt
\begin{tabular}{cccccccccc}
  \hline
  \hline
Reactions & $T/T_{\rm c} $ & $a_1$ & $b_1$ & $e_1$ & $a_2$ & $b_2$ & $e_2$ &
$d_0$ & $\sqrt{s_{\rm z}} $\\
\hline
 $I=\frac{1}{2}~\pi K^* \to \rho K$
  &  0     & 1.09 & 0.25  & 3.8  & 1.99 & 0.14  & 0.55  & 0.23 & 6.57\\
  &  0.65  & 0.37 & 0.03  & 0.52 & 0.68 & 0.1   & 0.48  & 0.05 & 4.02\\
  &  0.75  & 0.32 & 0.026 & 0.57 & 0.33 & 0.08  & 0.42  & 0.03 & 3.63\\
  &  0.85  & 0.18 & 0.017 & 0.62 & 0.24 & 0.047 & 0.41  & 0.02 & 2.9\\
  &  0.9   & 0.17 & 0.055 & 0.5  & 0.23 & 0.016 & 0.5   & 0.02 & 1.56\\
  &  0.95  & 0.15 & 0.071 & 0.64 & 0.29 & 0.02  & 0.47  & 0.03 & 1.35\\
  \hline
 $ I=\frac{1}{2}~\pi K^* \to \rho K^*$
  &  0     & 0.94  & 0.39  & 0.44 & 1.13  & 0.1   & 0.64 & 0.12 & 6.97\\
  &  0.65  & 0.183 & 0.1   & 0.5  & 0.151 & 0.5   & 0.8  & 0.15 & 4.26\\
  &  0.75  & 0.044 & 0.107 & 0.74 & 0.072 & 0.259 & 0.44 & 0.15 & 3.77\\
  &  0.85  & 0.017 & 0.05  & 0.55 & 0.038 & 0.14  & 0.46 & 0.08 & 2.48\\
  &  0.9   & 0.054 & 0.046 & 0.16 & 0.15  & 0.011 & 0.71 & 0.01 & 1.74\\
  &  0.95  & 0.25  & 0.005 & 0.29 & 1     & 0.021 & 0.74 & 0.02 & 1.11\\
  \hline
 $I=\frac{1}{2}~\pi K \to \rho K^*$
  &  0     & 0.34  & 0.95  & 0.76 & 0.5   & 0.19  & 0.51 & 0.25 & 9.63\\
  &  0.65  & 0.21  & 0.17  & 0.53 & 0.13  & 1.09  & 3.9  & 0.2  & 5.94\\
  &  0.75  & 0.051 & 1.34  & 6.8  & 0.102 & 0.22  & 0.5  & 0.2  & 4.85\\
  &  0.85  & 0.016 & 1.33  & 4.1  & 0.03  & 0.16  & 0.47 & 0.15 & 4.05\\
  &  0.9   & 0.014 & 0.52  & 0.51 & 0.02  & 0.06  & 0.45 & 0.1  & 3.78\\
  &  0.95  & 0.02  & 0.013 & 0.7  & 0.27  & 0.018 & 0.5  & 0.02 & 2.92\\
  \hline
 $ I=\frac{1}{2}~\rho K \to \rho K^*$
  &  0     & 1.59 & 0.08  & 0.76 & 1.69 & 0.26  & 0.4  & 0.1    & 5.02\\
  &  0.65  & 0.37 & 1.19  & 7.9  & 0.4  & 0.29  & 0.5  & 1.1    & 5.05\\
  &  0.75  & 0.2  & 0.25  & 0.48 & 0.39 & 1.31  & 6.9  & 1.35   & 4.48\\
  &  0.85  & 0.12 & 0.08  & 0.37 & 0.35 & 1.79  & 3.6  & 1.65   & 3.55\\
  &  0.9   & 0.58 & 0.01  & 0.32 & 0.538 & 1.85 & 15   & 1.85   & 3.38\\
  &  0.95  & 1.03 & 0.019 & 0.483 & 0.744 & 2.07 & 16  & 2.05   & 3.29\\
  \hline
  \hline
\end{tabular}
\end{table*}

\newpage
\begin{table*}[htbp]
\caption{\label{table3}Isospin-averaged unpolarized cross sections versus 
$T/T_{\rm c}$.}
\tabcolsep=5pt
\begin{tabular}{ccccccccc}
  \hline
  \hline
$\frac{T}{T_{\rm c}}$ & 0 & 0.6 & 0.65 & 0.68 & 0.7 & 0.72 & 0.75 & 0.78\\
$\sigma_{\pi \pi \to \rho}^{\rm un}$ (mb)
& 80.07 & 54.6  & 46.93 & 41.51 & 37.52 & 33.28 & 26.59 & 19.82\\
$\sigma_{\pi K \to K^*}^{\rm un}$ (mb)
& 60.5  & 42.53 & 38.79 & 34.97 & 31.27 & 26.79 & 19.66 & 15.39\\
  \hline
  \hline
$\frac{T}{T_{\rm c}}$ & 0.8 & 0.82 & 0.85 & 0.88 & 0.9 & 0.92 & 0.95 & 0.99\\
$\sigma_{\pi \pi \to \rho}^{\rm un}$ (mb)
& 15.94 & 13.98 & 12.37 & 11.43 & 10.93 & 10.18 & 7.62  & 0.85\\
$\sigma_{\pi K \to K^*}^{\rm un}$ (mb)
& 13.73 & 12.29 & 10.74 & 9.82  & 9.36  & 8.69  & 6.43  & 0.41\\
  \hline
  \hline
\end{tabular}
\end{table*}

\end{document}